\documentclass[pre,10pt,floatfix,superscriptaddress,showpacs,groupedaddress,nofootinbib,a4paper]{revtex4}
\usepackage{amsmath,bm,graphicx}
\usepackage{amssymb}
\usepackage{amsfonts}

\usepackage{placeins}
\usepackage{setspace}
\usepackage{soul}
\usepackage[dvips,usenames]{color}

\begin{document}

\title{Active colloids at fluid interfaces}

\author{P. Malgaretti}
\email[corresponding author: ]{malgaretti@is.mpg.de}
\author{M.N. Popescu}
\author{S. Dietrich}
\affiliation{Max-Planck-Institut f\"{u}r Intelligente Systeme, Heisenbergstr. 3, D-70569
Stuttgart, Germany}
\affiliation{IV. Institut f\"ur Theoretische Physik, Universit\"{a}t Stuttgart,
Pfaffenwaldring 57, D-70569 Stuttgart, Germany}

\date{\today}

\begin{abstract}
If an active Janus particle is trapped at the interface between a liquid and a fluid, 
its self-propelled motion along the interface is affected by a net torque on the 
particle due to the viscosity contrast between the two adjacent fluid phases. For a 
simple model of an active, spherical Janus colloid we analyze the conditions under which 
translation occurs along the interface and we provide estimates of the corresponding 
persistence length. We show that under certain conditions the persistence length of such 
a particle is significantly larger than the corresponding one in the bulk liquid, 
which is in line with the trends observed in recent experimental studies. 
\end{abstract}



\maketitle

\section{Introduction\label{Intro}}

Micro- and nanometer scale particles capable of self-induced motility within liquid 
environments \cite{ebbens,LaugaRev,SenRev,Sanchez2015,Poon} are promising
candidates for the development of novel lab-on-a-chip cell-sorting devices, chemical sensors
\cite{RevLabChip}, or targeted-drug-delivery systems \cite{RevDrugDelivery}, to cite just 
a few potential applications. One proposal, which has generated significant 
experimental and theoretical attention within the last decade (see, e.g., the recent reviews 
in Refs. \cite{SenRev,ebbens,Sanchez2015}), is to achieve self-motility by designing ``active'' 
particles capable to induce chemical reactions within the surrounding liquid. One such
system, which will be of particular interest for the present study, is represented 
by spherical beads partially covered over a spherical cap region by a catalyst which 
promotes, in the suspending solution, a chemical conversion of reactant 
(``fuel'') into product molecules\footnote{Such particles, which have distinct material 
properties across the two regions of their surface, are often called Janus 
particles; in the following we shall use this notion, too.}. Due to the partial 
coverage by catalyst, the spherical symmetry of the system is broken in two ways. First, 
the material properties of the particle vary across the surface and an axis of 
symmetry, which passes through the center of the particle and the pole inside the catalyst 
covered cap, can be defined. Second, the chemical reaction takes place only on the catalytic 
part of the surface, and therefore the chemical composition of the surrounding solution is 
varying along the surface of the particle. The out-of-equilibrium chemical composition 
gradients along the surface, due to the chemical reaction, couple to the particle via 
the interactions of the molecules in solution with the surface of the particle. This 
interplay eventually leads to hydrodynamic flows and to the motion of the particle 
relative to the solution, analogous to classic 
phoresis \cite{Anderson1989,Golestanian2005}. The active motion of such particles in the 
homogeneous bulk of the fluids has been the subject of numerous experimental (see, e.g., 
Refs. \cite{SenRev,ebbens,Golestanian2012,Fisher2014,Sanchez2015}) and theoretical (see,
e.g., Refs. \cite{Golestanian2005,Julicher,Popescu2009,Kapral2013,Lowen2011}) studies. 

However, in many cases such active Janus particles are suspended in a solution bounded by 
a liquid-fluid interface, which raises several new issues. It is known that in thermal 
equilibrium, i.e., in the \textit{absence} of such motility-promoting chemical reactions, 
owing to their amphiphilic nature Janus particles tend to accumulate at liquid-fluid 
interfaces. (This effect can be exploited, e.g., for the stabilization of 
binary emulsions \cite{Aveyard2012,Park2012,Rezvantalab2013}.) If the Janus particles 
are trapped at and confined to liquid-fluid interfaces their collective behavior, e.g., 
when externally driven or when relaxing towards equilibrium after a perturbation, can be 
strongly affected by this quasi two-dimensional ($2D$) confinement 
itself and by interface-promoted interactions, such as capillary interactions 
(see, e.g., Refs. \cite{alvaro2013,alvaro2011,alvaro2005}). 

Turning to the case of an \textit{active} Janus colloid, being trapped at the 
interface (i.e., being unable to move in the direction normal to the interface) 
can, on one hand, affect the particle dynamics due to the effects discussed above. On the 
other hand, this trapping may induce novel self-propulsion mechanisms. For example, 
it has been recently predicted that if one of the reaction products exhibits a 
preference for the surface and thus tends to accumulate at the interface, the trapped
Janus sphere will be set into motion along the interface by Marangoni flow, induced 
by the spatially non-uniform distribution of the reaction products 
\cite{Lauga2011,Stone2014}. (A similar motility mechanism can originate from 
thermally induced Marangoni flows, e.g., if the Janus particle contains a metal cap which 
is heated by a laser beam \cite{Wurger2014}; furthermore, as reported recently, 
induced Maragoni flows can drive the motion of active particles even if they are 
\emph{not trapped at} but \emph{located nearby} the 
interface~\cite{Dominguez2015}.)

If none of the reaction products exhibits a preference for the interface, the 
Marangoni type of propulsion is no longer in action. The question arises if for an active 
Janus particle, trapped at the interface, sustained motion along the 
interface can still occur due to the self-induced phoresis mechanism, which works 
in the bulk solution. For reasons of simplicity, in the following we focus solely on the case 
of planar interfaces. In thermal equilibrium (i.e., in the absence of diffusiophoresis), a 
Janus particle trapped at the interface typically exhibits a configuration in which the 
particle axis is not aligned with the normal of the interface (see Fig. \ref{fig:stab-janus}). 
Therefore, at first it seems that, upon ``turning on'' the chemical reaction, motion along the 
interface may be achieved\footnote{Even if the equilibrium configuration would correspond to
the symmetry axis being aligned with the interface normal, as, e.g., in the case 
of strongly repulsive, effective interactions between the interface and the catalytic 
patch, fluctuations will perturb this state of alignment and the previous scenario is 
recovered. These fluctuations can be thermal fluctuations of the orientation of the 
axis around the equilibrium position, or non-equilibrium fluctuations of the rate of the 
catalytic reaction along the surface.}. However, it is known that the motion at the 
interface between two fluids generally involves a coupling between translation and 
rotation \cite{LaugaRev,BrennerBook,Pozrikidis2007}. Thus the possibility arises that the 
translation along the interface may lead to a rotation of the axis of the particle 
towards alignment with the interface normal. Since self-phoresis of active particles 
is, in general, characterized by very small Reynolds (Re) numbers (i.e., inertia does not 
play a role) \cite{Anderson1989,ebbens,SenRev,Julicher}, such a rotation of the axis of 
the particle leads to a motionless state once the axis is aligned with the normal, i.e., 
the motile state is just a transient. Therefore, predicting whether or not for a particular 
system sustained motion along the interface may occur via self-diffusiophoresis requires an 
understanding of the interplay between the equilibrium configuration of the Janus particle 
at the interface, the distribution of reactant and product molecules upon turning on the 
chemical reaction, and the induced hydrodynamic flows in the liquid and the fluid.  

Here we study theoretically the issue of sustained self-diffusiophoresis along a 
liquid-fluid interface for a simple model of a spherical, chemically active Janus colloid 
trapped at a liquid-fluid interface. Nonetheless we expect this simple model to 
qualitatively capture some of the main physical features of the phenomenon. The chemical 
activity of the particle is modeled via the production of one species of solute molecules, 
with a uniform rate across the catalytic region. We determine the conditions under which 
this model system exhibits sustained motility. These conditions involve the interplay 
between the equilibrium configuration, the difference in viscosity between the two 
adjacent fluids, and the interactions between the particle and the product (solute) 
molecules. Finally, for particles trapped at the interface we analyze the persistence 
length and the stability of the motile state against thermal fluctuations and compare it 
with the corresponding motion in unbounded fluids. 

\section{Model\label{model}}

\begin{figure*}[!htb]
\includegraphics[width = 0.55\textwidth]{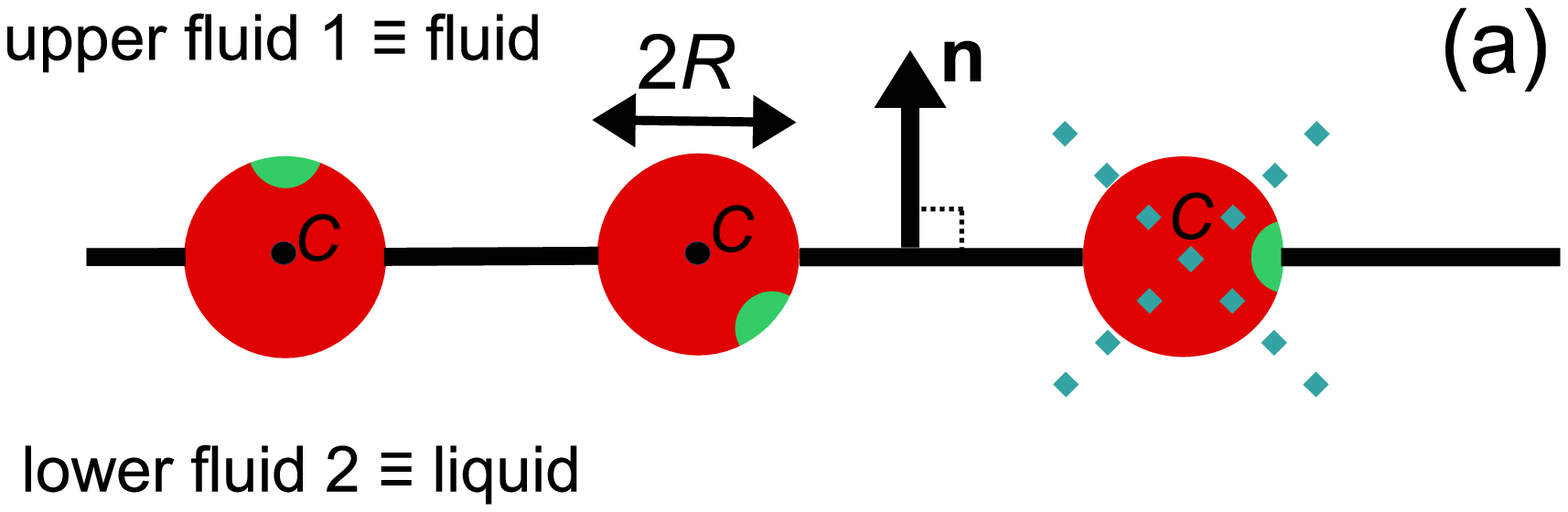}
\hspace*{0.03\textwidth}
\includegraphics[width = 0.4\textwidth]{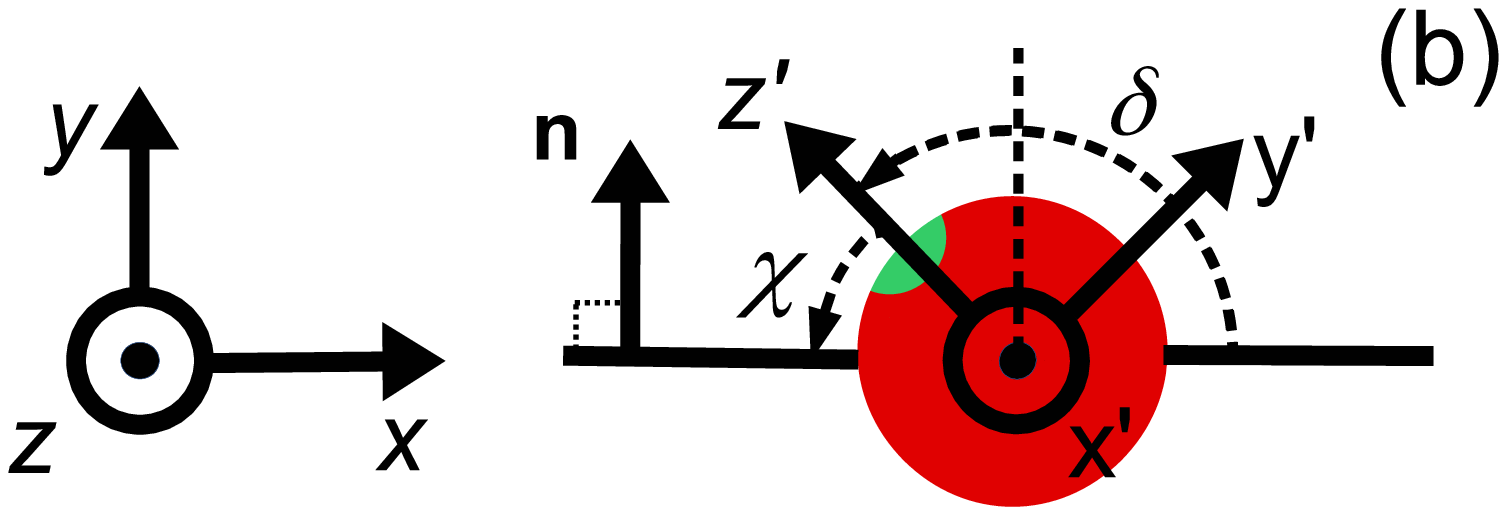}
\caption{
\label{fig:stab-janus}
(a) Typical configurations of an active Janus particle, i.e., a spherical colloid 
(shown as a red disk in sidewise projection onto the plane spanned by the symmetry 
axis of the particle and the interface normal $\mathbf{n}$) with a spherical cap 
providing a catalytic region (green), trapped at the interface (solid black line) 
between two fluids. The leftmost configuration corresponds to the catalytic cap having a 
preference for the upper fluid and very strong, effective repulsive interactions between the 
interface and the catalytic cap. The configuration in the middle corresponds to the 
catalytic cap exhibiting a preference for the lower fluid and effective interactions 
between the interface and the catalytic cap, having a long-ranged attractive 
component and a dominant short-ranged repulsive one. The configuration on the right 
(``crossed out``) does not occur due to the assumed preference of the catalyst for contact 
with one of the two fluids. In all three configurations $C$ denotes the center of the 
sphere. (b) The coordinate systems employed to describe the motion of such trapped active 
Janus colloids are shown at the right. In the spatially fixed unprimed coordinate system 
the interface normal $\mathbf{n}$ points into the $y$-direction, the $x$-direction is in 
the plane spanned by $\mathbf{n}$ and the symmetry axis of the particle, and the 
unit vector in the $z$-direction lies in the plane of the interface. The primed coordinate 
system is co-moving (translating and rotating) with the particle. The origin $O'$ of 
the primed coordinates system coincides with the center $C$ of the sphere. The $z'$- 
and $y'$- directions lie in the plane spanned by $\mathbf{n}$ and the symmetry axis 
of the particle. The $z'$-axis passes through the center of the catalytic cap, and the 
unit vector in the $x'$-direction lies in the plane of the interface. The angle $\delta 
\in [\delta_{m}, \pi - \delta_{m}) \cup [\pi + \delta_{m}, 2\pi - \delta_{m})$, 
where $\delta_{m} < \pi/2$ denotes the value of $\delta$ (with the orientation of 
the cap in the upper fluid, as shown in the figure) at which the rim of the cap touches 
the interface, gives the orientation of the $O'z'$ direction with respect to $Ox$, where 
$O$ is the origin of the fixed coordinate system. The unit vectors of the primed and 
unprimed coordinate systems are related via $\mathbf{e}_{x'} = \mathbf{e}_{z}$, 
$\mathbf{e}_{z'} = \cos(\delta) \mathbf{e}_{x} +  \sin(\delta) \mathbf{e}_{y}$, and 
$\mathbf{e}_{y'} = \sin(\delta) \mathbf{e}_{x} -  \cos(\delta) \mathbf{e}_{y}$. 
}
\end{figure*}

In this section we discuss our model for a catalytically active, spherical Janus
colloid \cite{Golestanian2005} trapped at a liquid-fluid interface (see Fig. 
\ref{fig:stab-janus}(a)). The colloid (red disk in Fig. \ref{fig:stab-janus}(a)) of radius 
$R$ has a spherical cap region (the green patch in Fig. \ref{fig:stab-janus}(a)) decorated 
by a catalyst which promotes the conversion $A \longrightarrow B$ of ``fuel'' molecules $A$ 
into product (solute) molecules $B$. The particle is trapped at the interface (the 
horizontal black line in Fig. \ref{fig:stab-janus}(a)) between the two fluids ``1'' and 
``2'' (denoted also as fluid and liquid, respectively) with bulk viscosities $\eta_1$
and $\eta_2$, respectively. For simplicity we assume 
that the fuel ($A$) and the product ($B$) molecules diffuse freely in both fluids, 
that neither of the two species $A$ and $B$ exhibits a preference for either the interface or 
 one of the two fluid phases, and that the concentration of $A$ molecules in the two fluid 
phases is at a steady state. Furthermore, it is assumed that the fuel molecules $A$ are 
present in abundance such that their number density is not affected by the reaction. 
(Under the latter assumption, the dynamics of the fuel molecules is irrelevant and their 
sole role is, similarly to that of the catalyst, that of a ``spectator'' enabling the 
reaction due to which active motion emerges. Thus within this model the diffusion 
constants of the $A$ particles in the two fluids do not enter the description.) 
Accordingly, we assume that the effects of the chemical reaction can be approximated by 
representing the catalyst area as an effective source of solute which releases molecules 
$B$ (at a time independent rate per area of catalyst). We denote the diffusion 
constants of the solute molecules in the two fluids by $D_1$ and $D_2$, respectively.

For a Janus particle trapped at the interface several scenarios can emerge if the parameters 
controlling  the effective particle-interface interactions, such as the coverage by 
catalyst or the three-phase contact angle of the ``bare'' particle (i.e., without 
catalyst) are changed. Here we restrict the discussion to the case that the particle and 
the two fluid phases have densities and surface tensions for which the deformations of the 
interface due to buoyancy are negligible, and we assume that the catalytic cap is 
completely immersed in one of the two fluid phases. Thus the Janus particle forms the 
three-phase contact angle of the bare particle \cite{Aveyard2012}. Furthermore, we assume 
that the effective interaction between the particle and the interface is such that the 
catalytic cap cannot jut into the other fluid, i.e., the circular boundary between the 
catalytic cap and the bare regions gets pinned at the three-phase contact line 
(between fluid 1, fluid 2, and the particle surface) upon touching it. (Accordingly, 
if the symmetry axis  of the particle, i.e., the axis passing through the center $C$ of 
the particle and the center of the catalytic spherical cap, would rotate beyond this 
touching point, the interface would no longer be planar.) For simplicity, we further 
restrict our discussion to the case in which the three-phase contact angle of the bare 
particle is $\pi/2$, which implies that the catalytic cap has to be smaller than a 
hemisphere. Actually, the size of the catalytic cap should be sufficiently small so that 
thermal fluctuations of the orientation around the equilibrium one do not lead to the 
aforementioned touching, which causes pinning of the liquid-fluid interface. With these 
assumptions, the center of the Janus particle lies in the plane of the interface. (As it 
will be discussed in Sec. \ref{sec:phoretic_velocities}, this particular configuration 
significantly simplifies the technical details, and thus provides transparent and 
physically intuitive results.) The orientation of the symmetry axis is determined by the 
effective interactions between the catalytic cap and the interface. For net repulsive 
interactions \cite{Aveyard2012} we expect the equilibrium distribution of the symmetry 
axis of the particle to be peaked at the direction normal to the interface. If, on the 
other hand, the effective interaction between the catalytic cap and the interface includes 
a long-ranged attractive part and a dominant short-ranged repulsive part, the equilibrium 
distribution is expected to be peaked at a direction which is close to, but distinct from, 
an orientation parallel to the interface. (Since the catalytic material is assumed to 
be completely immersed in one of the fluids, a net attractive interaction between the 
catalytic cap and the interface would be incompatible with our model.)

Since the surface tension compensates any action of the active Janus particle in the 
direction of the interface normal (which implies that the particle is trapped at the 
interface), translation of the particle upon turning on the catalytic reaction is possible 
only within the planar interface. Thus a motile state can be reached only if the axis of 
the Janus particle is not oriented perpendicular to the interface. Due to the symmetry of the 
problem, all lateral directions of the particle translation are equivalent; in other 
words, at a given tilting angle of the symmetry axis with respect to the normal, 
upon rotating the symmetry axis around the normal a state of motion emerges which is 
identical to the one in the original configuration. This allows us to consider the 
particle motion in the plane spanned by the axis of the particle (in its orientation at 
the moment when the catalytic reaction is turned on) and the normal of the planar interface. 
Thus we neglect the effects of thermal fluctuations leading to a rotation of the axis of 
symmetry out of this plane. Under this assumption, and in accordance with 
Fig. \ref{fig:stab-janus}(b), we choose the coordinate system with the $y$-axis along the 
interface normal, pointing towards  the upper fluid, the $x$-axis as the intersection of 
the interface with the plane of motion, and the $z$-axis as the normal of the latter 
(see Fig. \ref{fig:stab-janus}(b)). For future reference, we also introduce a system of 
coordinates -- with the origin $O'$ at the center $C$ of the particle and co-moving 
with the particle --, which is denoted by primed quantities. As shown in 
Fig. \ref{fig:stab-janus}(b), in the plane spanned by the axis of symmetry and the 
normal to the interface passing through the center of the particle we choose the $z'$-axis 
to point through the center of the catalytic cap and the $x'$-axis to lie in the plane 
of the interface and to be parallel to the $z$-axis. (These choices for the primed and 
unprimed coordinates are taken as to facilitate more convenient calculations in   
Sects. \ref{sec:phoretic_velocities} and \ref{sec:results} below.)

A viscosity contrast between the two fluids forming the interface leads to the onset 
of net torques on particles translating along the interface. Therefore the orientation of 
the symmetry axis of the trapped Janus particle relative to the $y$-axis will change 
once the chemical reaction is turned on and the particle is set into translation. 
Depending on the viscosity contrast between the two fluids, a small fluctuation of the 
orientation of the symmetry axis can be either amplified by the induced torque, leading 
to a different, yet motile, state, or suppressed. In the former case the steady state 
orientation of the axis of the Janus particle is ultimately determined by the 
geometry of the particle including the shape of the catalytic patch and the details 
of the effective interactions between the catalytic cap and the interface, which is a 
complex problem. Here we shall assume that the motion is quasi-adiabatic, in 
the sense that the rotation of the particle is much slower than the time it takes for 
the distribution of solute molecules $B$ and for the flow of the solution to reach a 
quasi-steady-state corresponding to the instantaneous orientation of the particle. 
Thus we focus solely on sustained motile states. The determination of the steady-state 
orientation of the symmetry axis of the particle (which, as noted above, ultimately 
involves the details of the effective interaction between the catalytic cap and the 
interface) is left to future research. 

As discussed above, the rotations (spinning) of the particle around the symmetry axis 
are neither contributing to, nor being induced by, the motility along the interface, whereas 
the rotations of the symmetry axis around the interface normal have the sole effect of 
changing the direction of the in-plane translation. Therefore describing the motion of the 
particle at the interface requires only to account for one translational velocity component  
$U_x$ and one angular velocity component $\Omega_z$ corresponding to 
translation along the $x$-axis and rotation around the $z$-axis, respectively. These will be 
determined by assuming the motion of the fluids to be described by the Stokes equations and 
the translation and rotation of the particle to be quasi-adiabatic in the sense that the 
hydrodynamics obeys the (steady state) Stokes equations at the instantaneous state, 
i.e., the orientation and velocity, of the particle. Finally, we assume that the interface 
exerts no force on the particle when it is translating along the interface, that the torques 
exerted by the interface with respect to rotations of the particle around the $z$-axis are 
vanishingly small, and that the eventual, small deformations of the interface at the contact 
line region accompanying such a motion also have negligibly small effects. 

The assumption of negligible torques deserves further consideration. In fact, if the 
catalytic cap leaves its equilibrium orientation, a torque will arise due to the 
effective interaction between the catalytic cap and the interface. Here we focus our 
attention on the case in which such contributions are negligible\footnote{Since the 
typical effective forces between the interface and the catalytic cap decay rapidly with 
the distance from the interface, the assumption remains valid as long as the catalytic cap 
is not very close to the interface.}. Furthermore, the rotation around the $x'$-axis 
involves a moving contact line, which is a well-known conceptual issue in classical 
hydrodynamics \cite{Huh1971,Dusan1974}. Here we do not attempt to provide a mechanism 
through which the  associated contact line singularity (i.e., translation of the contact 
line while at the same time the fluids fulfill the no-slip boundary condition) is removed 
and motion occurs (see, e.g., Refs. \cite{Dusan1976,Hocking1977,Brenner1986}). Instead we 
assume that the region, where a microscopic description is necessary, is very small 
compared to the typical length scales in the system and that there the expected 
macroscopic velocity values provide a smooth interpolation.

\section{Translational and angular velocities of Janus particles at liquid-fluid 
interfaces \label{sec:phoretic_velocities}}

In order to calculate the angular and translational velocity of Janus particles 
trapped at a liquid-fluid interface, we model the two immiscible fluids separated by a 
planar interface as having a continuous, but steeply varying, viscosity profile 
$\eta(y)$ interpolating between $\eta_2$ at $y=-\infty$ and $\eta_1$ at $y = +\infty$ 
across the plane $y = 0$ (compare Fig. \ref{fig:stab-janus}):
\begin{equation}
\eta(y) = \eta_{0} + \dfrac{\Delta\eta}{2} \tanh\left(\frac{y}{\xi}\right)\,,
\label{eq:def-visc-1}
\end{equation}
where $\eta_0 = (\eta_1+\eta_2)/2$ and $\Delta\eta = \eta_1 - \eta_2$ denote the 
mean viscosity $\eta_0 = \eta(0)$ and the viscosity contrast $\Delta \eta$, 
respectively, while $\xi$, which is of molecular size, characterizes the width of the 
interface. In the following we consider the case of a vanishingly thin interface, 
i.e., we take $\xi \to 0$.

\subsection{Reciprocal theorem for a particle at liquid-fluid interfaces}

We exploit the reciprocal theorem \cite{Stone1996,BrennerBook,Stone2014}, derived first 
by Lorentz \cite{Lorentz_original} (the English translation of the original paper is 
provided in Ref. \cite{Lorentz_transl}) for the case of a homogeneous fluid and later 
extended by Brenner \cite{BrennerBook} to the case in which the viscosity of the fluid 
varies spatially.\footnote{A recent extension of this version of the reciprocal theorem to 
the case of a free interface, in which the viscosity exhibits an abrupt change 
across the liquid-air interface, can be found in Ref. \cite{Stone2014}.} 
The reciprocal theorem states that in the absence of volume forces any two 
incompressible flow fields $\mathbf{u}(\mathbf{r})$ and 
$\hat{\mathbf{u}}(\mathbf{r})$, which are 
distinct solutions of the Stokes equations \textit{within the same domain $\cal D$}, 
i.e., solutions subject to different boundary conditions but \textit{on the 
very same boundaries $\partial \cal D$},  obey the relation
\begin{equation}
\int\limits_{\partial \cal D}^{} \mathbf{u} \cdot \hat{\boldsymbol{\sigma}} \cdot 
\mathbf{n} \,dS 
= \int\limits_{\partial \cal D}^{} \hat{\mathbf{u}} \cdot \boldsymbol{\sigma} \cdot 
\mathbf{n} \,dS\,,
\label{rec-theo-1}
\end{equation}
where $\boldsymbol{\sigma}$ and $\hat{\boldsymbol{\sigma}}$ denote the stress tensors 
corresponding to the two flow fields. 

For our system, the assumption of immiscibility of the two fluids translates into the 
kinematic boundary condition that the velocity components of the flow fields above 
and below the interface along the direction of the normal to the interface must vanish at the 
interface. This effectively enforces the interface as a physical boundary across 
which, concerning the hydrodynamics, there is momentum transfer but no mass 
transfer. Therefore it is necessary that the hydrodynamic flow is obtained by solving the 
Stokes equations in the domains above (${\cal D}_1$) and below (${\cal D}_2$) the interface 
and by subsequently working out the problem by connecting the solutions corresponding 
to the upper and lower fluids via appropriate boundary conditions at the interface 
(see below). ${\cal D}_1$ is delimited by that part ${\Sigma_p}_1$ of the 
particle surface ${\Sigma_p}$ exposed to the upper fluid, the surface of the fluid at 
infinity in the half-plane $y > 0$, and the upper part, $y = 0^+$ of the fluid 
interface $\Gamma$ (note that this is the plane $y = 0$ less the area occupied by 
the particle); ${\cal D}_2$ is defined similarly\footnote{Formally, the 
domains ${\cal D}_1$ and ${\cal D}_2$ as well as the interface $\Gamma$ could be closed 
along the $Ox$ direction by assuming periodic boundary conditions at $|x| \to \infty$; 
alternatively, one may choose for the surface at infinity, which is closing the 
domains ${\cal D}_1$ and ${\cal D}_2$, a spherical one, centered at $C$ and with a radius 
$R_\infty \to \infty$.}. We note that the inner normals of the upper ($\mathbf{n}_1$) 
and lower ($\mathbf{n}_2$) parts of the interface $\Gamma$ are 
$\mathbf{n}_1~= \mathbf{e}_y = -\mathbf{n}_2$. 

We restrict the discussion to the case in which for both flow fields (i.e., the 
un-hatted and the hatted one, which are to be considered in the reciprocal theorem), 
there are no (e.g., externally imposed, or due to surface tension gradients) tangential 
stresses at the interface. For both the un-hatted and the hatted flow fields, the 
corresponding flow velocities and stress tensors within the upper and lower domains, 
which are connected via the boundary conditions they have to obey at the interface, are 
denoted by the indices ``1'' and ``2'', respectively. By i) applying the 
reciprocal theorem (Eq. (\ref{rec-theo-1})) in each of the domains ${\cal D}_1$ and 
${\cal D}_2$ (which can be done because $\Gamma$ is a physical boundary at which the 
boundary conditions are formally prescribed by imposing the tangential velocity and stress 
tensor to take the values given by the (yet unknown) velocity and stress tensor on the 
other side of the interface, respectively); ii) adding the left and right hand sides of 
the two results of applying the reciprocal theorem in ${\cal D}_1$ and ${\cal D}_2$; iii) 
noting that for flow fields, which decay sufficiently fast with the distance from the 
particle (which typically is the case), the contribution from the integrals over the 
surfaces at infinity are vanishingly small; iv) using the relation (see above) 
$\mathbf{n}_1 = -\mathbf{n}_2 = \mathbf{e}_y$ between the interface normals; v) noting 
that the boundary conditions for the flow velocity at the interface $\Gamma$ impose zero 
normal components, i.e., $(\mathbf{u}_1 \cdot \mathbf{e}_y)|_{y = 0} = 0, (\mathbf{u}_2 
\cdot \mathbf{e}_y|)_{y = 0} = 0$ and continuous tangential components \cite{BrennerBook} 
so that $\mathbf{u}_1|_{y = 0_+} = \mathbf{u}_2|_{y = 0_-} =: 
u_{||} \mathbf{e}_{||}$ (with similar relations for the hatted velocity field), 
where $\mathbf{e}_{||} \cdot \mathbf{e}_y = 0$, we arrive at 
\begin{equation}
\label{rec-theo-interf} 
\int\limits_{\Gamma}^{} (u_{||} \mathbf{e}_{||}) \cdot ({\hat{\boldsymbol{\sigma}}}_1 - 
{\hat{\boldsymbol{\sigma}}}_2)|_{y = 0} \cdot \mathbf{e}_y \,dS + 
\int\limits_{\Sigma_p}^{} \mathbf{u} \cdot \hat{\boldsymbol{\sigma}} \cdot 
\mathbf{n} \,dS
= \int\limits_{\Gamma}^{} ({\hat{u}}_{||} \mathbf{e}_{||}) \cdot 
({\boldsymbol{\sigma}}_1 - 
{\boldsymbol{\sigma}}_2)|_{y = 0} \cdot \mathbf{e}_y \,dS + 
\int\limits_{\Sigma_p}^{} \hat{\mathbf{u}} \cdot \boldsymbol{\sigma} \cdot 
\mathbf{n} \,dS
\,.
\end{equation}
In the absence of tangential stresses at the interface the difference of 
normal stresses at the interface $({\boldsymbol{\sigma}}_1 - {\boldsymbol{\sigma}}_2)|_{y 
= 0} \cdot \mathbf{e}_y$ is the force corresponding to the Laplace pressure 
\cite{BrennerBook}, and thus is a vector oriented along the normal $\mathbf{e}_y$ of
the interface. Therefore the first integral on the left hand side of Eq. 
(\ref{rec-theo-interf}) vanishes due to $\mathbf{e}_{||}~\perp~\mathbf{e}_y$. By a 
similar argument, the first integral on the right hand side of Eq. 
(\ref{rec-theo-interf}) vanishes, too. Thus for the present system the reciprocal 
theorem takes the simple form given in Eq. (\ref{rec-theo-1}) but with $\cal D$ 
replaced by $\Sigma_p$.

In order to determine the translational and the angular velocity of the Janus particle, 
we shall select a proper set of ``dual problems'' (the ``hatted'' quantities), typically 
associated with known solutions for spatially uniform translations or rotations (under the 
action of external forces or torques) of solid spheres with prescribed boundary conditions 
at their surfaces. To this end we consider a solid sphere of radius $R$ with 
\textit{no-slip} boundary conditions translating with velocity $\hat{\mathbf{U}}$ and 
rotating with angular velocity $\hat{\mathbf{\Omega}}$ under the action of the external 
force $\hat{\mathbf{F}}$ and the external torque $\hat{\mathbf{L}}$. At the surface of 
the particle, the flow $\hat{\mathbf{u}}(\mathbf{r}_p)$, where $\mathbf{r}_p$ denotes a 
point at the particle surface $\Sigma_p$, is given by
\begin{equation}
\hat{\mathbf{u}}(\mathbf{r}_p) = \hat{\mathbf{U}} + \hat{\mathbf{\Omega}} \times 
(\mathbf{r}_p - \mathbf{r}_{\mathrm{c}}) \,,
\label{eq:surf_flow_reciproc}
\end{equation}
where $\mathbf{r}_\mathrm{C}$ is the position of the center $C$ of the sphere. (Both 
$\mathbf{r}_p$ and $\mathbf{r}_\mathrm{C}$ are measured from a common, arbitrary origin, 
the location of which drops out from $\mathbf{r}_p - \mathbf{r}_\mathrm{C}$.) Similarly, 
we consider a Janus particle, which translates with velocity $\mathbf{U} = 
U_x \mathbf{e}_x$ and rotates with an angular velocity 
$\mathbf{\Omega}=\Omega_z\mathbf{e}_z$ around the axis, which is parallel to $Oz$ and 
passes through the moving center $C$ of the particle at its instantaneous position. If 
the particle exhibits boundary conditions given by a \textit{phoretic slip} velocity 
$\mathbf{v}(\mathbf{r}_p)$, the flow $\mathbf{u}(\mathbf{r}_p)$ at its surface is 
given by
\begin{equation}
\mathbf{u}(\mathbf{r}_p) = \mathbf{U} + \mathbf{\Omega}\times(\mathbf{r}_p - 
\mathbf{r}_\mathrm{C}) + \mathbf{v}(\mathbf{r}_p) \,.
\label{eq:surf_flow_Janus}
\end{equation}
By using Eq. (\ref{eq:surf_flow_reciproc}) and noting that $\hat{\mathbf{U}}$ and 
$\hat{\mathbf{\Omega}}$ are spatially constant vectors, the right hand side (rhs) of 
Eq. (\ref{rec-theo-1}) can be re-written as
\begin{equation}
\int\limits_{\Sigma_p}^{} \hat{\mathbf{u}} \cdot \boldsymbol{\sigma} \cdot \mathbf{n} \,dS 
= \hat{\mathbf{U}} \cdot \mathbf{F} + \hat{\mathbf{\Omega}} \cdot \mathbf{L}\,,
\label{rhs_rec_theo}
\end{equation}
where $\mathbf{F} = \int_{\Sigma_p} \boldsymbol{\sigma} \cdot \mathbf{n}\, dS$ and 
$\mathbf{L} = \int_{\Sigma_p} (\mathbf{r}_p-\mathbf{r}_\mathrm{C}) \times 
\boldsymbol{\sigma} \cdot \mathbf{n}\, dS$ denote the force and the torque, respectively, 
experienced by the Janus particle. (In Eq. (\ref{rhs_rec_theo}), $\mathbf{n}$ denotes the 
normal of the surface $\Sigma_p$ of the particle, oriented into the fluid.) We note that 
while the active motion of Janus particles in the bulk is force- and torque-free, this is, 
in general, not the case if the motion occurs at the interface because the interface can 
exert forces and torques on the particle. However, because we have assumed that the 
interface does not exert a force in the case of translations of the Janus particle along 
the interface or a torque in the case of rotations around the $z$-axis (in the sense of 
spinning around an axis parallel to the $z$-axis, as discussed above), the components 
$F_x$ and $L_z$ vanish. Therefore, if the reciprocal problem involves only translations 
along the $x$-axis and/or such rotations around the $z$-axis, the rhs of Eq. 
(\ref{rhs_rec_theo}), and, consequently, of Eq. (\ref{rec-theo-1}) is zero. Restricting 
now the dual problem as discussed above to such a choice, and using Eq. 
(\ref{eq:surf_flow_Janus}) for the left hand side of Eq. (\ref{rec-theo-1}), we arrive 
at 
\begin{equation}
U_x \hat{F}_x + \Omega_z \hat{L}_z = -\int\limits_{\Sigma_p}^{} \mathbf{v}(\mathbf{r}_p) 
\cdot \hat{\boldsymbol{\sigma}} \cdot \mathbf{n} \,dS \\.
\label{eq:rec-theo-4}
\end{equation}

\subsection{Calculation of the translational and angular velocities}
We proceed by selecting two so-called dual problems, each involving only one of the 
two types of motion (translation or rotation only) for both of which Eq. 
(\ref{eq:rec-theo-4}) holds. These will provide two relations allowing one to 
determine $U_x$ and $\Omega_z$. The first one, denoted by the index ``1'', is that of a 
sphere of radius $R$, the center of which lies in the plane of the flat, sharp 
($\xi\rightarrow 0$) liquid-fluid interface (see Eq. (\ref{eq:def-visc-1})), 
translating \textit{without rotation} with velocity $\hat{\mathbf{U}} = \hat U_x 
\mathbf{e}_x$ along the interface. This problem has been 
solved analytically \cite{Pozrikidis2007}, and the result of interest here is
\begin{equation}
\left.(\mathbf{n}\cdot\hat{\boldsymbol{\sigma}}_1)\right|_{\Sigma_p}=
-\frac{3}{2R}\eta(\mathbf{r}_p)\hat{\mathbf{U}}\,.
\label{trasl-norot-noslip}
\end{equation}
This leads to (see Appendix A for the details of the calculation)
\begin{subequations}
\label{eq:recipr-prob-1}
\begin{equation}
\label{F1x}
\hat{F}_{1x} =  -6 \pi R \,\eta_{0} \,\hat{U}_x := \alpha_1 \hat{U}_x
\end{equation}
and
\begin{equation}
\label{L1z}
\hat{L}_{1z} = +\frac{3 \pi}{2} R^{2} \Delta\eta \, \hat{U}_x := \beta_1 \hat{U}_x\,.
\end{equation}
\end{subequations}
After inserting Eqs. (\ref{trasl-norot-noslip}) and (\ref{eq:recipr-prob-1}) into 
Eq. (\ref{eq:rec-theo-4}) and canceling the common factor $\hat {U}_x$, we obtain
\begin{equation}
\alpha_1 U_x + \beta_1 \Omega_z = C_1:= \frac{3}{2R} 
\int\limits_{\Sigma_p}^{} \eta(\mathbf{r}_p) \, v_{x}(\mathbf{r}_p) \,dS\,,
\label{eq:rec-ther-1}
\end{equation}
where $v_x(\mathbf{r}_p)$ is the $x$-component of the phoretic slip velocity at the 
point $\mathbf{r}_p$ on the surface of the particle.

The second problem which we consider, denoted by the index ``2'', is that of 
the driven rotation, \textit{without translation}, with angular velocity 
$\hat{\mathbf{\Omega}} = \hat{\Omega}_z \, \mathbf{e}_z$ of a spherical particle of radius 
$R$, the center of which lies in the plane of the flat, sharp ($\xi\rightarrow 0$) 
liquid-fluid interface. As noted before, this is a more involved problem due to the 
concomitant issue of contact line motion. For the case in which one of the two fluids has 
a vanishingly small viscosity, an exact solution was constructed in 
Ref. \cite{Brenner1986} under the assumption that a slip boundary condition, with a 
spatially uniform slip length along the surface, applies across that surface region 
which is immersed in the fluid of non-vanishing viscosity, called liquid. The 
result of this calculation shows that for typical slip lengths $l_0$, which are much 
smaller than the size of the particle,\footnote{For example, for water on PDMS or on 
glass surfaces the estimated slip length $l_0$ is well below 100 nm 
\cite{Joseph2005}.} at distances $|y|/l_0\gg 1$ from the interface the hydrodynamic 
flow within the liquid is \textit{de facto} identical with the one which would 
have occurred if the rotating sphere would have been completely immersed in the 
liquid and a no-slip boundary condition would have been applied. Thus in this context 
the only role played by the slip is to remove the contact line singularity, as 
discussed in Sect. \ref{model}. 
This can be interpreted in the sense that the same solution would emerge if one 
assumes that the fluid slips only in a narrow region localized close to the 
three-phase contact line, while the no-slip condition holds for the rest of the 
sphere. This view is confirmed by an alternative solution presented in Ref. 
\cite{Sterr2009} for the same problem of the rotation of a sphere at the
interface between a liquid and a fluid of vanishing viscosity.

In the following we shall adopt the latter interpretation and make the 
\textit{ansatz} that for our above problem ``2'' (in which the viscosities of 
both fluids are, in general, certain non-zero quantities) with a sharp 
interface ($\xi\rightarrow 0$) the expression for the stress tensor at the surface of the 
particle is given by the one in Ref. \cite{Brenner1986}, i.e., 
\begin{equation}
\left.(\mathbf{n}\cdot\hat{\boldsymbol{\sigma}}_2)\right|_{\Sigma_p} = -3 \,
\eta(\mathbf{r}_p) \,\hat{\Omega}_z \, 
\left.(\mathbf{e}_z\times \mathbf{n})\right|_{\Sigma_p}\,,
\label{notrasl-rot-noslip}
\end{equation}
except for a small region localized close to the three-phase contact line. As noted, 
the expression above is expected to provide a reliable approximation if the viscosity 
contrast between the two fluids is large \cite{Brenner1986}. In the limiting case 
of the two fluids becoming identical, by construction Eq. (\ref{notrasl-rot-noslip}) 
reduces to the exact result corresponding to a sphere rotating without slip in a 
spatially homogeneous fluid.

With the assumption that the small region near the three-phase contact line 
contributes negligibly to the integrals over the surface of the particle, Eq. 
(\ref{notrasl-rot-noslip}) implies that the corresponding components of the forces and 
torques required for the reciprocal theorem (Eq. (\ref{eq:rec-theo-4})) are given by 
(see Appendix A)
\begin{subequations}
 \label{eq:recipr-prob-2}
\begin{equation}
\label{F2x}
\hat{F}_{2x}  \simeq  +3 \pi R^{2} \Delta\eta \,\hat{\Omega}_z
:= \alpha_2 R \,\hat{\Omega}_z
\end{equation}
and
\begin{equation}
\label{L2z}
\hat{L}_{2z} \simeq  -8 \pi \eta_{0} R^{3} \hat{\Omega}_z
:= \beta_2 R \,\hat{\Omega}_z\,.
\end{equation}
\end{subequations}
After inserting Eqs. (\ref{notrasl-rot-noslip}) and (\ref{eq:recipr-prob-2}) into 
Eq. (\ref{eq:rec-theo-4}) and canceling the common factor $\hat {\Omega}_z$, we obtain
\begin{equation}
\alpha_2 U_x + \beta_2 \Omega_z = C_2 := \dfrac{3}{R} \int\limits_{\Sigma_p}^{} 
\eta(\mathbf{r}_p) \left(\mathbf{n} \times \mathbf{v}(\mathbf{r}_p) \right)_{z}\,dS\,.
\label{eq:rec-ther-2}
\end{equation}

Once a particular model is given for the mechanism through which the chemical 
activity determines the phoretic slip $\mathbf{v}(\mathbf{r}_p)$, the quantities $C_1$ 
and $C_2$ can be computed and $U_x$ and $\Omega_z$ follow from Eqs. (\ref{eq:rec-ther-1}) 
and (\ref{eq:rec-ther-2}). This concludes the calculation of the translational $(U_x)$ and 
angular $(\Omega_z)$ velocities of the Janus particle trapped at the interface. We note that 
in the limit $\Delta\eta \rightarrow 0$ Eqs. (\ref{eq:rec-ther-1}) and (\ref{eq:rec-ther-2}) 
reduce to the corresponding components for a Janus particle moving in a homogeneous fluid 
\cite{Stone1996}, i.e., $U_x = - \frac{1}{4 \pi R^{2}} \int\limits_{\Sigma_p}^{} v_{x} dS$ 
and $\Omega_z = -\frac{3}{8\pi R^{3}} \int\limits_{\Sigma_p}^{} \left(\mathbf{n} \times 
\mathbf{v}\right)_{z}dS$, respectively. 

This latter result deserves further consideration. In the case of translation along 
the interface, in the limit $\Delta \eta\rightarrow 0$ the recovery of the result 
corresponding to a particle moving in a homogeneous fluid is to be expected in the case of 
a particle \textit{having its center located at the interface}. This is so, because
the flow around the Janus particle translating at the interface converges, as the viscosity 
contrast approaches zero, towards the solution corresponding to the motion in a homogeneous 
fluid \footnote{In the case of a homogeneous fluid the flow around the particle which 
translates is symmetric with respect to any plane containing the translation direction. 
Thus the flow is characterized by a vanishing velocity normal to such a symmetry 
plane and a continuous velocity tangential to that symmetry plane (which thus 
can be regarded as an  ``imaginary planar interface'' where the kinematic boundary 
conditions of an actual interface between immiscible liquids are obeyed).}. On the other 
hand, this expectation does not hold in the case of rotation: the immiscibility of the 
fluids requires that the interface modifies the flow by ``forcing'' the fluid to 
flow along the interface. This different structure of the flow survives even 
if the viscosity contrast is vanishing. Therefore, recovering nonetheless the 
result for rotation in a homogeneous fluid simply means that the \textit{ansatz} for the 
stress tensor \textit{at the surface of the particle} (Eq. (\ref{notrasl-rot-noslip})) 
renders the correct limiting behavior for vanishing viscosity contrast, irrespective of 
the corrections provided by the presence of the interface.

\section{Results and discussion \label{sec:results}}

Solving Eqs. (\ref{eq:rec-ther-1}) and (\ref{eq:rec-ther-2}) for $U_x$ and $\Omega_z$ 
leads to
\begin{subequations}
\label{eq:solu-tot-1}
\begin{equation}
\label{Ux}
U_x  =  \dfrac{C_1 \beta_2 - \beta_1 C_2}{\alpha_1 \beta_2 - \alpha_2 \beta_1} 
\end{equation}
and
\begin{equation}
\label{Omz}
\Omega_z  =  \dfrac{\alpha_1 C_2 - C_1 \alpha_2}{\alpha_1 \beta_2 - \alpha_2 \beta_1}\,.
\end{equation}
\end{subequations}
For the discussion of these results we find it more convenient to employ the following 
alternative description of the orientation of the particle with respect to the 
interface. We define the director $\mathbf{p}$ of the Janus particle as being the unit vector 
corresponding to the axis of symmetry of the Janus particle oriented towards the catalytic cap. 
The acute angle between the director and the interface, in the plane $(xOy)$, is denoted by 
$\chi$ (Fig. \ref{fig:stability-condition}(a)). It is defined as a signed 
quantity, with the sign convention that $\chi$ is positive if $\mathbf{p}$ points 
towards the  half-space $y > 0$ and negative otherwise; thus $-\pi/2 \leq \chi 
\leq \pi/2$. (The angle $\chi$ is thus connected with $\delta$ (see Fig. 
\ref{fig:stab-janus}(b)) via $\chi = \mathrm{min}(\delta,\pi-\delta)$, if $0 < \delta 
< \pi$, and $\chi = - \mathrm{min}(\delta-\pi,2 \pi-\delta)$, if $\pi < \delta < 2 
\pi$.) The states with $\chi = \pm \,\pi/2$ correspond to the director being 
parallel and antiparallel, respectively, to the interface normal $\mathbf{e}_y$, for which 
$U_x$ vanishes. Therefore, in order for a state of motion along the surface, i.e., $U_x 
\neq 0$, to be sustainable, any change in $\chi$ occurring as a result of motion 
along the interface should be such that $|\chi|$ decreases (i.e., the director 
rotates towards the interface).

\subsection{Configurations of sustained motility}

\begin{figure*}[!htb]
\includegraphics[width = 0.22\textwidth]{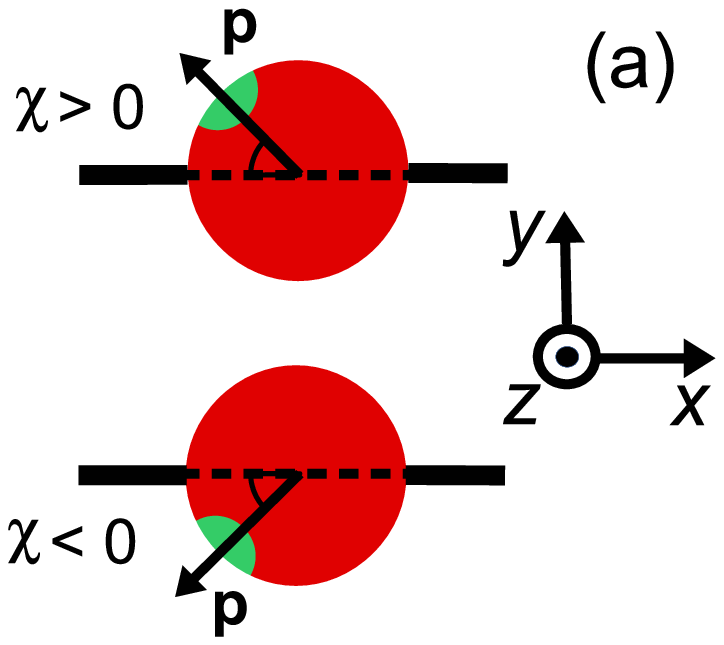}
\hspace*{0.05\textwidth}
\includegraphics[width = 0.69\textwidth]{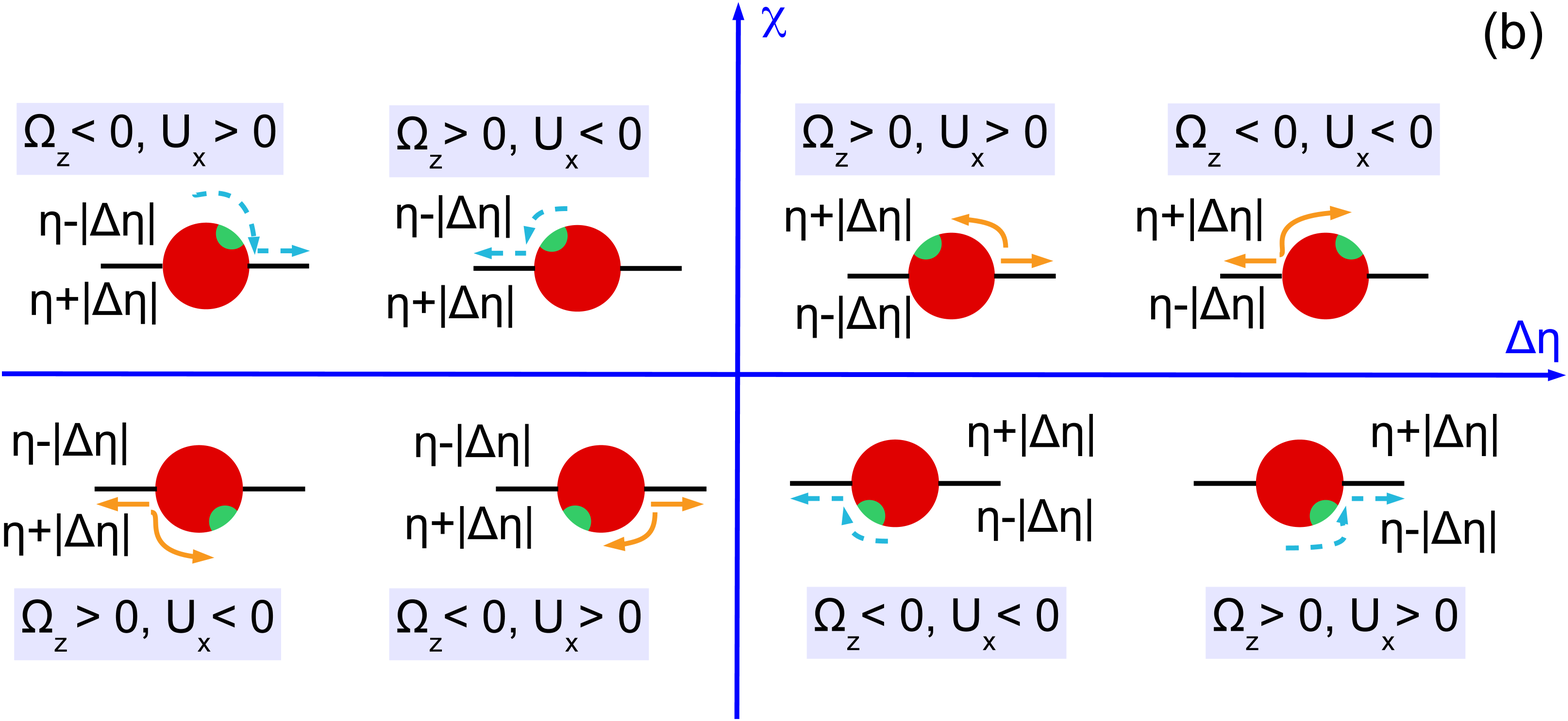}
\caption{
\label{fig:stability-condition}
(a) Definition of the director orientation with respect to the interface. 
(b) Schematic diagram indicating the configurations of sustained motility as function of 
the viscosity contrast $\Delta\eta = \eta_1-\eta_2$ and the position of the active 
cap (green) with respect to the interface (black) given by the acute angle $\chi$ 
between the director and the interface. The orange and blue colors of the arrows refer to 
repulsive and attractive interactions between the chemically generated solute and the 
particle, respectively. In each case we have illustrated both the 
situation in which the catalytic cap is tilted \textit{to the left} and the situation 
in which it is tilted \textit{to the right} of the interface normal $\mathbf{e}_y$, 
respectively. This change in the tilting of the cap amounts to all blue and 
orange arrows turning around and pointing into opposite directions but without changing 
colors because the stability criterion ($\mathrm{sign}(U_x/(R \Omega_z))$) is 
invariant with respect to this change. Therefore in each quadrant only either 
the repulsive or the attractive interaction case is associated with sustained motion.
}
\end{figure*}

Referring now to Fig. \ref{fig:stability-condition}(b) and considering as an 
example the situation shown in the upper part with the catalytic cap tilted slightly 
to the left of the normal $\mathbf{e}_y$, for which $\chi > 0$, one infers that, 
upon turning on the chemical reaction, for repulsive (attractive) interactions between 
the solute (i.e., the reaction products) and the particle the latter will tend to move 
towards the right (left), so that  $U_x > 0$ ($U_x < 0$). If $\Delta \eta > 0$, 
i.e., the upper fluid is more viscous than the lower one, translation with $U_x > 0$ gives 
rise to a torque on the particle which induces a counterclockwise rotation, i.e., 
$\Omega_z > 0$. (The upper part of the particle experiences a stronger, retarding 
friction than the lower part of the particle.) This corresponds to a decrease of 
$\chi$ towards zero and thus promotes motility. This situation is shown in the right 
upper quadrant of Fig. \ref{fig:stability-condition}(b). On the other hand, 
translation with $U_x < 0$ (and still for $\chi >0$ as well as the cap tilted to the left 
of the interface normal; not shown in the right upper quadrant of Fig. 
\ref{fig:stability-condition}(b)) gives rise to a clockwise rotation (i.e., $\Omega_z < 
0$, for the same reason as above) and therefore to an increase of $\chi$ towards 
$\pi/2$, i.e., rotation opposes motility. If $\Delta \eta < 0$ (and still $\chi > 0$ 
with the cap tilted to the left of the interface normal), the sign of those torques 
(which are described by the same color), and thus of the corresponding angular velocities, 
is reversed. In this case, the translation towards the left ($U_x < 0$) is accompanied by 
a rotation which decreases $\chi$ (i.e., $\Omega_z > 0$) and thus promotes motility 
(see the left upper quadrant of Fig. \ref{fig:stability-condition}(b)), while translation 
towards the right (i.e., $U_x > 0$ and still $\chi > 0$ with the cap tilted to 
the left of the interface normal; not shown in the left upper quadrant) is opposed by the 
rotation of the director. Following the above reasoning for the various possible 
configurations (i.e., catalytic cap above or below the interface, attractive or repulsive 
solute-particle interactions, viscosity contrast positive or negative), in the plane 
$(\Delta \eta, \chi)$ one can identify the cases in which sustained motion would occur, 
depending on the repulsive or attractive character of the interactions between the solute 
and the particle. These configurations are summarized in Fig. 
\ref{fig:stability-condition}(b), where the arrows indicate the corresponding directions 
of the translation and rotation. The colors blue and orange of the arrows refer to 
repulsive and attractive interactions, respectively.

From the discussion above (see also the schematic diagram in Fig. 
\ref{fig:stability-condition} (b)) one infers that the states with sustained motion 
must satisfy $U_x/(R \Omega_z) > 0$ for $\Delta \eta > 0$ or $U_x/(R \Omega_z) < 0$ for 
$\Delta \eta < 0$. Therefore, for a given system these signs of $U_x/(R \Omega_z)$ as a 
function of the viscosity contrast $\Delta \eta$ provide necessary conditions for the 
occurrence of such motile states. However, in order to explicitly calculate the sign 
of the ratio $U_x/(\Omega_z R)$ one needs to provide an explicit form for 
$\mathbf{v_s}(\mathbf{r}_p)$ which determines $U_x$ and $\Omega_z$ (Eqs. 
(\ref{eq:rec-ther-1}), (\ref{eq:rec-ther-2}), and (\ref{eq:solu-tot-1})). In order 
to determine $\mathbf{v_s}(\mathbf{r}_p)$ it is in principle necessary (i) to 
specify the geometrical properties of the catalytic cap responsible for the reaction 
within the fluids; (ii) to specify the reaction; (iii) to provide the diffusion 
constants of the reactant and product molecules in the two fluids (for example, they can
either diffuse in both fluids but with different diffusion constants, or some 
of the reactants or products may effectively be confined to one of the two fluids), as 
well as any effective interaction between these molecules and the interface (e.g., 
whether or not they act as surfactants); (iv) to provide the interaction 
potentials of the various molecular species in the two solutions (i.e., the two 
fluids plus the reactants and the products) with the Janus particle as a whole 
as well as with its surface (for both the catalyst covered part and the inert part).

\subsection{Motility of a model chemically active Janus colloid at fluid interfaces}

Here we focus on the simple model of a chemically active Janus particle as introduced in 
Sec. \ref{model}, for which there is only a single reaction product (``solute'') diffusing 
in both fluids and for which the reactant molecules are present in abundance and diffusing 
very fast in both fluids, such that in both fluids the number density of reactant 
molecules is \textit{de facto} time-independent and spatially uniform. The 
effective interaction of the solute with the colloidal particle is assumed to be 
of a range which is much smaller than the radius $R$ of the colloid and  to be similar for 
the catalyst-covered part and the inert part of the colloid. Furthermore, we assume a sharp 
interface (i.e., $\xi \to 0$ so that $\eta(\mathbf{r}_p) = \eta(y)$ with $\eta(y) = 
\eta_1$ for $y > 0$, while $\eta(y) = \eta_2$ for $y < 0$). The latter assumptions imply 
that, by adopting the classical theory of phoresis 
\cite{Anderson1989,Golestanian2005,Popescu2009} which has been developed for 
homogeneous (i.e., constant viscosity) fluids, one can express the phoretic slip as 
being proportional to the solute concentration gradient along the surface at all 
points of the surface of the particle except for a small region near the interface. 
Within the corresponding proportionality factor $\mathcal{L}/(\beta\,\eta)$ 
(the so-called ``phoretic mobility''; see, c.f., Eq.(\ref{def:slip_vel})), where 
$\beta = 1/(k_B T)$, $k_B$ is the Boltzmann constant, and $T$ denotes the absolute temperature, it is possible 
to identify the contribution $\mathcal{U}(h)$ of the solute-particle interaction 
(relative to the solvent-particle interaction). $\mathcal{U}(h)$ is encoded in 
$\mathcal{L}$ (which has the units of an area) according to \cite{Anderson1989}
\begin{equation}
 \mathcal{L}=\int\limits_{0}^{\infty} dh \, h 
 \left(e^{-\beta \mathcal{U}(h)} - 1 \right)\,,
 \label{eq:phor-mob}
\end{equation}
where $h$ is the distance between the point-like solute and the particle 
surface. The potential $\mathcal{U}(h)$ is assumed to be such that $\mathcal{U}(h 
\to 0) = + \infty$, i.e., right at the particle surface the solvent is strongly 
preferred. The potential can be either repulsive at all distances, or it can become 
attractive beyond a certain distance $h_0$ (and thus has to have an attractive minimum 
because at large distances it decays to zero); this latter case corresponds to 
\textit{adsorption} of the solute. Note that $\mathcal{L} < 0$ for purely repulsive 
interactions $\mathcal{U}(h)$, while if $\mathcal{U}(h)$ has an attractive part and $h_0$ 
is sufficiently small, one has $\mathcal{L} > 0$. In the following, the notion of
``attractive interactions'' will refer strictly to the latter case, i.e.,  potentials 
$\mathcal{U}(h)$ which have attractive parts and satisfy $\mathcal{L} > 0$. At this stage 
we do not yet particularize the cap to more than the assumed spherical cap shape and to 
being completely immersed into one of the two fluids. 

Under the above assumptions, the phoretic slip $\mathbf{v}(\mathbf{r}_p)$ follows 
from the solute distribution around the surface of the Janus particle. We further 
assume that the diffusivity $D(\mathbf{r})$ of the solute molecules is sufficiently 
high such that the number density distribution $\rho(\mathbf{r},t)$ of the 
solute is not affected by the convection of the fluids (i.e., we assume that 
the P\'eclet number $Pe$ is small) and that a steady state distribution 
$\rho(\mathbf{r})$ of solute is established at time scales which are much shorter than the 
characteristic translation time $R/U_x$ of the colloid. With this, $\rho(\mathbf{r})$ obeys 
the diffusion equation
\begin{equation}
 \nabla\cdot[D\nabla\rho]=0
 \label{eq:diff-eq}
\end{equation}
subject to the boundary conditions
\begin{equation}
 \lim_{r\rightarrow\infty}\rho=0,\,\,\,\,\, -D\nabla \rho\cdot\mathbf{n}|_{r=R} = Q \,
H(\theta_0 - \theta),
 \label{eq:diff-eq-bond-cond}
\end{equation}
where $\mathbf{n}$ is the outer normal of the particle, $Q$ denotes the number of solute 
molecules generated per area and per time at the location of the catalytic cap, and 
$H(x)$ is the Heaviside step function ($H(x > 0) = 1$, $H(x < 0) = 0$). (In accordance 
with the assumptions of the model (see Sec. \ref{model}), in the above diffusion 
equation there are no terms to account for eventual interactions of the solute with the 
interface or with external fields.)

Since actually only the distribution of solute at the particle surface is required in 
order to calculate the phoretic slip, instead of seeking for the full solution 
$\rho(\mathbf{r})$ of Eq. (\ref{eq:diff-eq}), which is a difficult problem, we only 
focus on the solute distribution at the particle surface. In the co-moving (primed) 
coordinate system (see Fig. \ref{fig:stab-janus}(b)), in which the phoretic slip 
velocity is most conveniently calculated, we introduce, in the usual manner, the 
common spherical coordinates $(r',\theta',\phi')$ defined via
\begin{equation}
x' = r' \sin\left(\theta'\right)\cos\left(\phi'\right)\,,~
y' = r' \sin\left(\theta'\right)\sin\left(\phi'\right)\,,~
z' = r' \cos\left(\theta'\right)\,.
\label{spherical-coord}
\end{equation}
Accordingly, the solute distribution at the particle surface $\rho(\mathbf{r'}_p) = 
\rho(R,\theta',\phi')$ can be expressed as a series expansion in terms of the spherical 
harmonics $Y_{\ell\,m} (\theta',\phi')$ 
\cite{Jackson_book}:
\begin{equation}
\rho(R,\theta',\phi')= \sum\limits_{\ell = 0}^\infty  
\sum\limits_{m = -\ell}^{m = \ell} A_{\ell,m}\, Y_{\ell\,m}(\theta',\phi') \,,
\label{def:rho_sph_harm}
\end{equation}
where the coefficients $A_{\ell,m}$ are functions of the radius $R$ and 
of the other parameters (temperature, diffusion constants, viscosities, rate of 
solute production, etc.) characterizing the system. 
In the co-moving system, the phoretic slip can be expressed in terms of the gradients 
of $\rho(\mathbf{r}'_p)$ along the surface of the particle 
\cite{Anderson1989,Golestanian2005,Popescu2009,Poon} (\textit{except} at the three-phase 
contact line):
\begin{equation}
\mathbf{v} (\mathbf{r}'_p)  = - \dfrac{\mathcal{L}}{\beta\,\eta(\mathbf{r}_p)} 
\nabla'_{||} \rho(\mathbf{r}'_p)\,:= v_{\theta'}\mathbf{e}_{\theta'} 
+v_{\phi'}\mathbf{e}_{\phi'}
\label{def:slip_vel}
\end{equation}
where $\nabla'_{||}=\frac{1}{R}\mathbf{e}_{\theta'}\partial_{\theta'}+
\frac{1}{R\sin\theta'}\mathbf{e}_{\phi'}\partial_{\phi'}$ denotes the projection of the 
gradient operator along the surface of the particle. 

In order to determine the $x$-component $v_x$ of the slip-velocity in the spatially fixed 
coordinate system, we use Eqs. (\ref{def:rho_sph_harm}) and (\ref{def:slip_vel}) and 
employ the relation between the unit vectors of the spatially fixed coordinate system and 
the co-moving one (see Fig. \ref{fig:stab-janus}(b)). Knowledge of $v_x$ allows one to 
determine the quantities $C_1$ and $C_2$ introduced in Eqs. (\ref{eq:rec-ther-1}) and 
(\ref{eq:rec-ther-2}) (see Appendix \ref{deriv-C1-C2} for details): 
\begin{equation}
 C_1 = -2 \sqrt{3\pi} \, \dfrac{\mathcal{L} \cos(\delta)}{\beta} A_{1,0} \,, 
\label{eq:C1B}
\end{equation}
and
\begin{equation}
 C_2 = 0 \,. \label{eq:C2B}
 \end{equation}
It is interesting to note that, apart from materials properties ($\cal L$) and 
temperature, $C_1$ depends solely on the projection ($\cos(\delta)$) of the 
particle director onto the plane of the interface and on the real amplitude 
$A_{1,0}$ (see Eq. (\ref{def:rho_sph_harm})) of $Y_{1\,0}(\theta',\phi') = 
\sqrt{3/(4 \pi)} \, \cos(\theta')$. This is that contribution to the angular 
dependence of $\rho$ along the particle surface which varies slowest between the 
poles at $\theta' = 0$ (center of the cap) and $\theta' = \pi$. This can be interpreted 
as an indication that for the model considered here the difference in the solute 
density between that at the catalytic pole and at the inert antipole is the dominant 
characteristics while the details of the variation of the density along the surface 
between these two values are basically irrelevant for the motion of the particle. 

In the case that the two fluids have the same viscosity, i.e., $\Delta \eta = 0$, and 
the diffusion constant for the product molecules is the same in the two fluids 
(e.g., being related to the viscosity via the Stokes-Einstein relation), for the model 
considered here, according to which the reactant and product molecules can 
diffuse freely in both fluids and unhindered by the interface, the diffusion 
equation (Eqs. (\ref{eq:diff-eq}) and (\ref{eq:diff-eq-bond-cond})) becomes identical to 
the one in a homogeneous bulk fluid which can be solved analytically 
\cite{Golestanian2005,Popescu2009}. (Thus, in this limit there is no signature of the 
interface left in the diffusion problem.) The corresponding expansion into spherical 
harmonics of the solute density at the surface of the particle (for a 
\textit{b}ulk solvent without an interface) leads to the expression 
\begin{equation}
\label{eq:hom_A10}
A_{1,0}^{(b)} = \varkappa \dfrac{Q R}{D_0}\,,
\end{equation}
where $\varkappa$ is a dimensionless factor determined by the geometry of the Janus 
sphere (i.e., the extent of the catalyst covered area) and $D_0$ is the diffusion 
constant of the product molecules in the fluids of viscosities $\eta_1 = \eta_2 = 
\eta_0$. This leads to the ansatz
\begin{equation}
\label{eq:def_varsig}
 A_{1,0} = \varsigma(\Delta\eta/\eta_0) A_{1,0}^{(b)}\,,
\end{equation}
for a system with an interface, where the dimensionless function $\varsigma$ is expected to 
depend on the viscosities solely via the dimensionless ratio $\epsilon = \Delta\eta/\eta_0$. 
Since in the limit $\epsilon \to 0$, which renders the homogeneous bulk fluid case, 
one has $A_{1,0} \to A_{1,0}^{(b)}$, the function $\varsigma$ must obey the constraint 
$\varsigma(\epsilon \to 0) = 1$.

By combining Eqs. (\ref{eq:recipr-prob-1}), (\ref{eq:recipr-prob-2}),
(\ref{eq:solu-tot-1}), (\ref{eq:C1B}), (\ref{eq:C2B}), 
(\ref{eq:hom_A10}), and (\ref{eq:def_varsig}), we obtain 
\begin{subequations}
\label{eq:solu-tot-2-explicit}
\begin{equation}
\label{Ux_expl}
U_{x}  = \dfrac{1}{\sqrt{3 \pi}} 
\dfrac{\varsigma\left(\dfrac{\Delta\eta}{\eta_0}\right)}
{1-\dfrac{3}{32}\left(\dfrac{\Delta\eta}{\eta_0}\right)^2} \cos(\delta) V_0  + {\cal 
O} \left(\left(\dfrac{\Delta \eta}{\eta_0}\right)^3\right)
\,,
\end{equation}
\begin{equation}
\label{Omz_expl}
\Omega_{z} = \dfrac{3}{8} \dfrac{1}{R} \dfrac{\Delta\eta}{\eta_0} \, U_x + {\cal 
O} \left(\left(\dfrac{\Delta \eta}{\eta_0}\right)^3\right)
\,,
\end{equation}
\end{subequations}
where 
\begin{equation}
 V_0 := \dfrac{\mathcal{L} \,A_{1,0}^{(b)}}{\beta \eta_0 R} 
= \varkappa \dfrac{\mathcal{L} Q}{\beta \eta_0 D_0}\,
\label{eq:def_V0}
\end{equation}
renders the characteristic translational and angular velocity scales $|V_0|$ and 
$\Omega_0 = |V_0|/R$, respectively. $V_0$ is independent of the particle radius $R$ 
as well as of the value of the viscosity $\eta_0$ because, under the assumption of the 
Stokes-Einstein relation, $\beta \eta_0 D_0$ depends only on the radius $R_m$ of the 
product \textit{m}olecules. This implies that the translational velocity is 
independent of the radius $R$ of the 
particle while the angular velocity is proportional to $1/R$ (up to eventual 
additional dependences on $R$ arising from $\varsigma$).

Equation (\ref{eq:solu-tot-2-explicit}) shows that both the translational and the angular 
velocity are proportional to $\cos\left(\delta\right)$. Therefore both vanish for 
$\delta=\pi/2$ which matches with the fact in this case the particle is 
in fully upright orientation and thus cannot propel laterally. Moreover, both $U_x$ 
and $\Omega_z$ change sign when the director $\mathbf{p}$ (Fig. 
\ref{fig:stability-condition}(a)) changes from pointing mainly to the 
right to pointing mainly to the left (Fig. 
\ref{fig:stability-condition}(b)). On the other hand, the sign of the ratio $U_x
/\Omega_z$, which, according to the discussion of Fig. \ref{fig:stability-condition}(b) 
in the main text, decides on the sustainability of the motile state, is independent 
of $\delta$ but is determined by the sign of $\Delta\eta$. This is in 
agreement with the symmetry exhibited by the diagram shown in Fig. 
\ref{fig:stability-condition}(b). In the limit of a vanishing viscosity 
contrast $\Delta\eta/\eta_0$, $U_{x}$ approaches the constant value $V_0 
\cos(\delta)/\sqrt{3 \pi}$. (This corresponds to the motion in a homogeneous 
bulk fluid under the constraint of moving along a plane at an angle $\delta$ with respect 
to the orientation of the director position.) Thus, for small values of $\Delta \eta$, 
$U_x(\Delta \eta)$ does not vary much, while, as expected, the angular 
velocity vanishes linearly $\propto \Delta\eta/\eta_0$. In the limiting case 
$\Delta\eta \to 0$ translation and rotation are decoupled and the particle translates 
without any rotation because there is no viscosity contrast. In such a case the net 
effect of the interface is to keep the particle center bound to the plane of the 
interface.

While the diagram in Fig. \ref{fig:stability-condition} is entirely determined by the 
ratio $U_x/\Omega_z$, which is independent of $\varsigma$, the magnitudes of both 
$U_x$ and $\Omega_z$ do depend on it via the amplitude $A_{1,0}$ (Eqs. 
(\ref{eq:solu-tot-2-explicit}) and (\ref{eq:def_V0})). Since determining the exact form of 
$\varsigma(\epsilon)$ is clearly analytically intractable, one can try to analyze its 
behavior for $\epsilon \ll 1$. One option is to employ a perturbation series in terms of 
the small parameter $\epsilon$ in order to calculate the distribution of solute for 
$\epsilon \ll 1$, starting from the known solution $\rho_0(\mathbf{r})$ for $\epsilon 
= 0$ (i.e., for a homogeneous bulk fluid without interface), from which one can estimate 
$A_{1,0}$ and implicitly $\varsigma(\epsilon)$. (Note that $\rho_0(\mathbf{r})$ varies 
spatially due to the solute sources located at the surface of the particle and the solute 
sink at infinity.) We denote by $\tilde{\rho}(\mathbf{r}) := \rho(\mathbf{r}) - 
\rho_0(\mathbf{r})$ and $\tilde{D}(\mathbf{r}) := D(\mathbf{r})-D_0$ the 
deviations (first order in $\epsilon$) of the number density distribution and of the 
diffusion coefficient from their corresponding values $\rho_0(\mathbf{r})$ and $D_0$ 
(spatially constant) in a homogeneous medium. (Note that by assuming the Stokes-Einstein 
relation between the diffusion coefficient and the viscosity $\tilde{D}(\mathbf{r})$ is 
a known function determined by $D_0$, $\epsilon$, and the known variation of the 
viscosity across the interface (Eq. (\ref{eq:def-visc-1})).) From Eqs. 
(\ref{eq:diff-eq}) and (\ref{eq:diff-eq-bond-cond}) one obtains that $\tilde{\rho}$ is the 
solution of the differential equation
\begin{equation}
\nabla \cdot \left(\tilde{D} \nabla\rho_0  + 
D_0 \nabla \tilde{\rho}\right)=0\,,
\label{eq:diff-1-order-a}
\end{equation}
subject to the boundary conditions
\begin{equation}
\lim_{r\rightarrow\infty}\tilde\rho=0,\,\,\,\,\,\, \left(\tilde D\nabla 
\rho_0+D_0\nabla \tilde{\rho}\right)\cdot\mathbf{n}|_{r=R}=0\,.
\label{eq:diff-1-order-c}
\end{equation}
We have been unable to find an analytical solution of Eqs.~(\ref{eq:diff-1-order-a}) and 
(\ref{eq:diff-1-order-c}) for a general orientation of the (small) cap. Therefore we 
cannot make any further rigorous statements. Instead, we only formulate 
expectations concerning the behavior of $\varsigma(\epsilon)$. For example, considering 
the case in which the catalytic cap is in the upper fluid region ($y > 0$), $\epsilon > 
0$ (i.e., enhanced [reduced] viscosity in the upper [lower] fluid) leads to a 
reduction [increase] of the diffusion coefficient in the upper [lower] fluid. Compared 
with the homogeneous fluid ($\epsilon=0$), intuitively this should lead to a relative 
accumulation of product molecules near the catalytic pole (located in the upper fluid) 
and to a relative depletion near the inert antipole (which is located in the lower 
fluid). For $\epsilon<0$ the behavior is reversed. Since (as discussed after Eq. 
(\ref{eq:C2B})) the coefficient $A_{1,0}$ can be viewed as a measure of the 
difference between the densities at the catalytic pole and at the antipole, the 
reasoning above suggests that, upon deviating from the homogeneous state (with 
$A_{1,0}^{(b)}$), $A_{1,0}$ varies oppositely if the viscosity of fluid ``1'' relative 
to that of fluid ``2'' increases or decreases, respectively. Therefore, to first 
order in $\epsilon$ the function $\zeta(\epsilon)$ is expected to vary as 
$\zeta(\epsilon\rightarrow 0)=1+\text{const}\cdot\epsilon+\mathcal{O}(\epsilon^2)$.

\subsection{Persistence length and effective diffusion coefficient for a 
chemically active Janus colloid at fluid interfaces}\label{sec:persitence length}

The motion of active particles is characterized by distinct regimes occurring at  
different time scales. At short time scales the active motion amounts to a ballistic 
trajectory whereas at larger time scales the behavior is diffusive. A key parameter, 
which characterizes the motion of active particles, is the persistence length 
(which can be defined as below irrespective of whether the active particle is trapped 
at an interface or moving in a bulk fluid)
\begin{equation}
 \lambda=\bar{v}\tau\,.
 \label{def:lambda}
\end{equation}
This is the typical distance a Janus particle, moving at an instantaneous velocity 
$\bar{v}$, covers before thermal fluctuations will eventually change its direction. The 
time $\tau$ is determined by the rotational diffusion of the particle 
(see Ref. \cite{GolestanianPRL2007}). In the present case of 
the active particle being trapped at the interface there are two types of rotations. 
First, there are rotations of the catalytic cap orientation, i.e., of 
$\mathbf{p}$ around the interface normal, with a 
characteristic time $\tau_\parallel$. These rotations lead $\mathbf{p}$ out of the 
initial plane of motion spanned by $\mathbf{p}$ and the interface normal. This clearly 
changes the direction of motion. Second, there are fluctuations of $\mathbf{p}$ within the 
plane of motion with the normal of the plane of motion acting as the rotation axis. 
Small fluctuations of this kind do not change the direction of motion because $\mathbf{p}$ 
stays within the initial plane of motion. However, large fluctuations can rotate 
$\mathbf{p}$, within the plane of motion, from a predominantly forward direction to a 
predominantly backward direction so that the particle runs backwards along the same 
straight line. This flipping of directions is associated with a time scale 
$\tau_\perp$. The minimum of these two time scales sets the rotational 
diffusion time $\tau_i = \min(\tau_\perp,\tau_\parallel)$ for an active 
particle trapped at an interface.

In the absence of thermal fluctuation the distribution function of the orientation 
of the axis $\mathbf{p}$ of the particle is peaked at the steady state value. 
Thermal fluctuations promote a broadening of the distribution. Both cases of 
rotations translate into fluctuations of the value of the instantaneous velocity of 
the particle. Accordingly, the typical velocity $\bar{v}_i$ of an active 
particle trapped at an \emph{i}nterface is defined as the mean velocity of the 
Janus particle obtained as a weighted integral over all those possible configurations 
which give rise to a velocity with the same prescribed sign\footnote{Since in 
the present case the system does not undergo any spontaneous symmetry breaking, the 
velocity obtained by averaging over \textit{all} possible configurations, rather than 
only over those with a prescribed sign of the velocity, is zero.}. Before entering 
into further technical details concerning the definition of $\bar{v}_i$, it is 
convenient to focus on one of the eight cases shown in Fig.~\ref{fig:stability-condition}(b), 
namely the case of a Janus particle characterized by $V_0 < 0$ (see Eq. (\ref{eq:def_V0}) 
for repulsive solute-particle interactions so that $\mathcal{L} < 0$) and with 
the catalytic cap in the \emph{upper} phase with $\delta < \pi/2$ ( so that 
$\chi = \delta > 0$). For $\Delta\eta > 0$, and within the linear regime 
$\epsilon \ll 1$, Eq. (\ref{eq:solu-tot-2-explicit}) renders, in this case, $U_x < 0$ 
and $\Omega_z < 0$. This is the situation illustrated in the right part of the top 
right quadrant of Fig.~\ref{fig:stability-condition}(b). The other cases can be discussed 
along the same line. Accordingly, we define 
$\bar{v}_i$ as
\begin{equation}
\bar{v}_i = \left|\, \int\limits_{\delta_m}^{\pi/2} \mathcal{P} 
(\delta) U_x(\delta)d\delta \right|,
 \label{def:bar-v}
\end{equation}
where $\mathcal{P}(\delta)$ is the steady state probability distribution to find a Janus 
particle with its axis forming an angle $\delta\in (\delta_m,\pi/2)$ with the plane of 
the interface\footnote{The mean velocity $\bar{v}_i$ for the same particle moving 
in the positive direction would be $\bar{v}_i = \int_{\pi/2}^{\pi-\delta_m}
\mathcal{P}(\delta) \,U_x(\delta) \,d\delta$; see the left part of the top right 
quadrant of Fig.~\ref{fig:stability-condition}(b). Here $\mathcal{P}$ is the distribution 
of the angle $\delta\in (\pi/2,\pi-\delta_m)$.}; $\delta_m$ is the value of $\delta$ for 
which the catalytic cap would touch the interface (see Fig.~\ref{fig:stab-janus}(b)). 

In thermal equilibrium, $\mathcal{P}(\delta)=\mathcal{P}_{eq}(\delta)$ depends only 
on the effective interactions between the catalytic cap and the interface. For 
example, in the absence of such interactions one has $\mathcal{P}_{eq}(\delta) = 
(\pi/2-\delta_m)^{-1}$. In contrast, in the presence of an effective attraction we expect 
that $\mathcal{P}_{eq}(\delta)$ exhibits a peak closer to the interface (i.e., close to 
$\delta = 0$) while the opposite holds in the case of an effective repulsion for which one 
expects $\mathcal{P}_{eq}(\delta)$ to be peaked  at $\delta = \pi/2$. When particles are 
active, an additional torque arises due to the catalytic activity, hence modifying 
the shape of $\mathcal{P}(\delta)$. In order to estimate $\mathcal{P}(\delta)$ for active 
particles, we assume that $\mathcal{P}(\delta)$ factorizes as 
$\mathcal{P}(\delta)=\mathcal{P}_{eq}(\delta)\mathcal{P}_{neq}(\delta)$ into an 
equilibrium part $\mathcal{P}_{eq}(\delta)$ (as discussed above) and a modulation 
$\mathcal{P}_{neq}(\delta)$ due to the particle activity. As far as 
$\mathcal{P}_{neq}(\delta)$ is concerned, we assume that, although the system is 
out of thermal equilibrium, we can express it as a Boltzmann weight 
$\mathcal{P}_{neq}(\delta)\propto e^{-\beta\Phi(\delta)}$ of an effective potential 
$\Phi(\delta)$ accounting for the torque arising due to the particle activity. Since 
within the present model there are no effective interactions with the interface, 
$\mathcal{P}_{eq}(\delta)$ is a constant which can be absorbed into the 
normalization:
\begin{equation}
 \mathcal{P}(\delta):= \mathcal{P}_{eq}(\delta) \mathcal{P}_{neq}(\delta) = 
\dfrac{e^{-\beta\Phi(\delta)}}{Z}
 \label{def:distr-prob}
\end{equation}
where $Z = \int_{\delta_m}^{\pi/2}e^{-\beta\Phi(\delta)}d\delta$ ensures that 
$\int_{\delta_m}^{\pi/2} \mathcal{P}(\delta)d\delta = 1$.

An estimate of the potential $\Phi$ can be obtained as follows. According to Eq. 
(\ref{eq:Lz_expr}), in our model a Janus particle, trapped at the interface and spinning 
with angular velocity $\Omega_z$ around an axis which is contained in the plane of the 
interface and passes through the center of the particle, experiences a torque 
$L_z = -8 \pi \eta_0 R^3 \Omega_z$. Therefore, in order to maintain the angular 
velocity $\Omega_z(\delta)$ of the particle an external torque equal to $-L_z$ must be 
applied to the particle. Accordingly, for the Janus particle translating with 
$U_x(\delta)$ while simultaneously rotating with $\Omega_z(\delta)$, we introduce an 
effective torque
\begin{equation}
L(\delta) = 8 \pi \eta_0 \Omega_z(\delta) R^3
\end{equation}
analogous to the external one which would have accounted for the same angular 
velocity, and define, via $L(\delta) = - d\Phi(\delta)/d\delta$, the effective potential 
$\Phi(\delta)$ as
\begin{equation}
\beta\Phi(\delta) :=- \beta\int\limits^{\delta}_{\delta_{m}} L(\delta')d\delta' 
~~\stackrel{Eqs. (\ref{eq:solu-tot-2-explicit}),\,(\ref{eq:def_V0})}{=} 
\Pi \,[ \sin(\delta_m)- \sin(\delta) ]\,.
\label{def:pot-PHI}
\end{equation}
In this equation one has $\Pi = \sqrt{3 \pi} \beta V_0 R^2 \Delta\eta$, the reference 
potential is set to $\Phi(\delta_m) = 0$, and we have accounted for the fact that here we 
discuss only the case in which $\delta_m \leq \delta \leq \pi/2$ (each of the other 
three quadrants can be analyzed following the same line of reasoning).

The sign of $\Pi$ is determined by the sign of $V_0$ and $\Delta\eta$. As we noted 
above, here we focus on the case in which $V_0<0$ (i.e., there is a repulsive 
interaction between the Janus particle and the product molecules of the catalysis) so 
that  $\Pi = -\sqrt{3 \pi}\, \beta |V_0| R^2 \Delta\eta = - 1/(2 \sqrt{3 \pi})
(\Delta\eta/\eta_0) Pe_0$, where $Pe_0 = |V_0| R/D_P > 0$ is the P\'eclet number of a 
Janus particle in a homogeneous fluid of viscosity $\eta_0$ and 
$D_P = k_B T /(6 \pi \eta_0R)$ is the diffusion constant of the Janus particle defined 
via the Stokes-Einstein relation. The above expression for $\Pi$ shows that, for 
$\Delta\eta > 0$ and a Janus particle characterized by $V_0 < 0$, $\Pi$ is negative. 
In this case one has $\Phi(\delta > \delta_m) > 0$ (Eq. (\ref{def:pot-PHI})) and 
thus $\Phi(\delta)$ attains its minimum at $\delta = \delta_m$. This means that the 
action of the effective torque is consistent with $\Omega_z < 0$ 
(see Eq. (\ref{Omz_expl})) which ``drives'' the particle towards its steady-state 
orientation $\delta_m$, as discussed in Fig.~\ref{fig:stability-condition}(b) 
(right part of the top right quadrant). Accordingly, in the right part of the 
top right quadrant, corresponding to $\Delta \eta > 0$ and to the director pointing 
into the upper fluid, consistent with $\delta_m \leq \delta < \pi/2$ and thus 
$\chi = \mathrm{min}(\delta,\pi-\delta) > 0$, repulsive interactions (orange arrows) 
ensure a sustained motility state by providing a torque which tilts the director 
towards the interface, i.e., which in the present case leads to a decrease 
of $\delta$ towards  $\delta_m$. As one can read off from Eq. (\ref{def:pot-PHI}), 
the characteristics of the Janus particle (see Eq. (\ref{eq:def_V0})) and of the 
fluid phases are all encoded in $\Pi$. Therefore the above conclusions can be extended 
directly to the case of attractive interactions between the Janus particle and the 
product molecules of the catalytic reaction by changing the sign of $V_0$, and hence 
of $\Pi$. Therefore, in the case of attractive interactions (i.e., $V_0 > 0 $) and with 
the cap oriented such that $\delta_m \leq \delta <\pi/2$ and $\chi = \delta > 0$, 
$\Pi$ is negative for $\Delta \eta < 0$ and thus $\Phi(\delta > \delta_m) > 0$ (in 
agreement with the situation illustrated in the left part of the top left 
quadrant of Fig.~\ref{fig:stability-condition}(b) for which $U_x > 0$ and $\Omega_z 
< 0$). Therefore, if particles have the catalytic cap in the \textit{upper} phase, for 
which $\sin(\delta) > 0$, the states with small values of $\sin(\delta)$, i.e., with 
the catalytic patch being closer to the interface and thus promoting the motile 
state, are favored if $\Pi < 0$. Similarly, if the catalytic cap of the particle is in 
the \textit{lower} phase, where $\sin(\delta) < 0$, large values of $\sin(\delta)$, i.e., 
the catalytic patch being closer to the interface and thus promoting the motile 
state, are favored if $\Pi > 0$. 

Concerning the persistence length $\lambda_i = \bar{v}_i \tau_i$ of an active 
particle trapped at an interface (Eqs. (\ref{def:lambda}) and (\ref{def:bar-v})) one 
would like to understand its relation to the persistence length $\lambda_b = \bar{v}_b 
\tau_b$ of a similar active particle moving freely in a homogeneous bulk fluid. To this 
end, we proceed by assuming that the characteristic time $\tau_i$ for the loss of 
orientation of a Janus particle, trapped at and moving along an interface between two 
fluids characterized by a (not too large) viscosity contrast $\Delta \eta \neq 0$, is 
similar to the corresponding characteristic rotational diffusion time $\tau_b$ for the 
loss of orientation in a homogeneous bulk fluid\footnote{It is 
particularly difficult to determine $\tau_i$ because it depends on the details of the 
effective interaction between the Janus particle and the interface and it involves the 
dynamics of the moving three-phase contact line.} of viscosity $\eta_0$. However, since 
one of the three possible independent rotations of a rigid body would affect the 
directionality for the particle trapped at the interface only if it is associated with a 
large fluctuation which would flip the director $\mathbf{p}$ with respect to the interface 
normal (see the discussion of $\tau_{\perp}$ after Eq. (\ref{def:lambda})), it is a 
reasonable to expect that $\tau_i > \tau_b$. Furthermore, the argument concerning the 
weak influence of the rotations associated with $\tau_{\perp}$ suggests $\tau_i = \nu_0 
\tau_b$ with $\nu_0 \simeq 3/2$ as a good \textit{ansatz} for the relation between the 
two characteristic time scales. Furthermore, we note that Eq. (\ref{Ux_expl}) has the 
form of a projection onto the $x$-axis (due to the factor $\cos(\delta)$) 
of a velocity (the factor multiplying $\cos(\delta)$) oriented along the director 
$\mathbf{p}$. Thus in the limit $\Delta \eta \to 0$ this latter factor can be 
identified with the velocity ${\bar v}_b$ of the active particle moving in a homogeneous 
bulk fluid of viscosity $\eta_0$. With $\varsigma(\epsilon \to 0) = 1$ this renders 
${\bar v}_b = \frac{1}{\sqrt{3\pi}} |V_0|$.

\begin{figure}
 \includegraphics[width = 0.47 \textwidth]{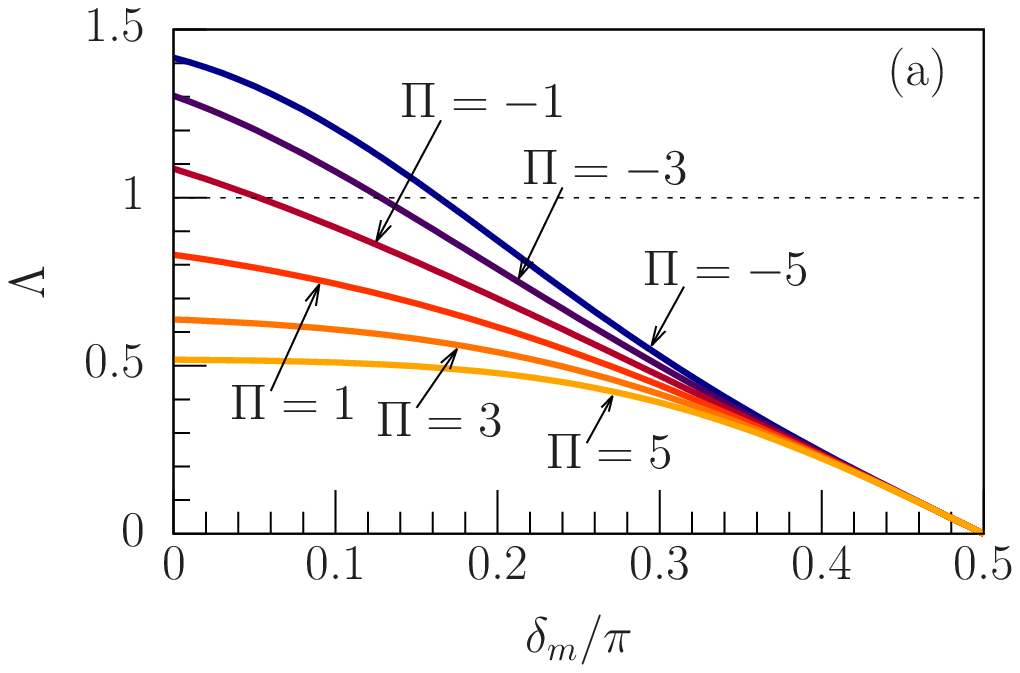}
  \includegraphics[width = 0.47 \textwidth]{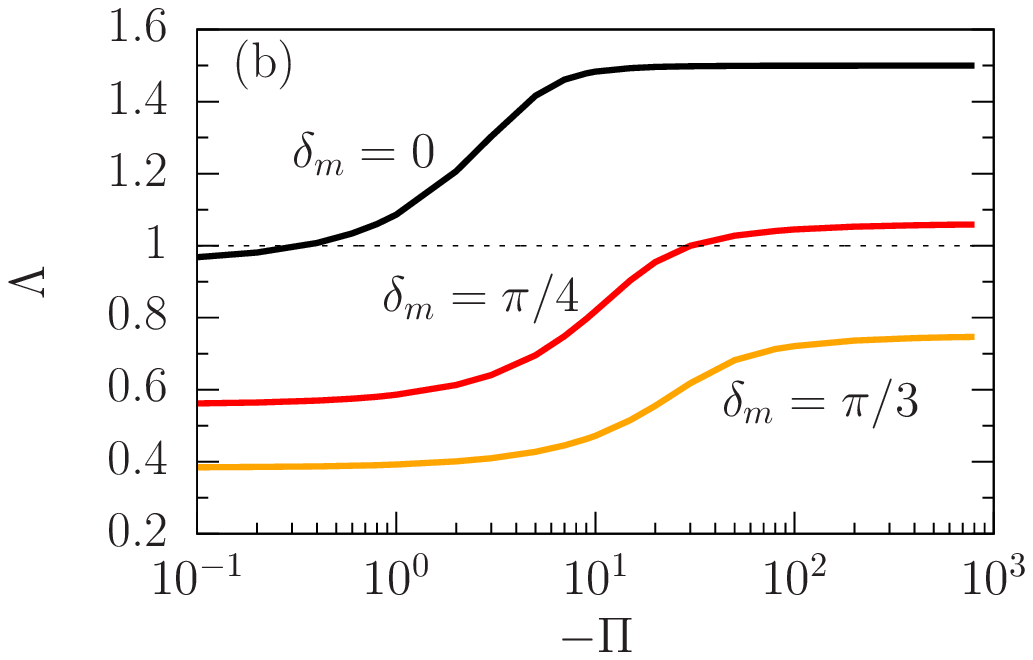}
\caption{
(a) Approximate expression for the ratio $\Lambda=\lambda_i/\lambda_b$ 
(Eq. (\ref{eq:Lamb_calc}) with $\nu_0 = 3/2$) of the persistence length of a Janus 
particle at the interface ($\lambda_i$, catalytic cap in the upper phase) and in the bulk 
($\lambda_b$) as a function of $\delta_m$ for various values of $\Pi = \sqrt{3 \pi} \beta 
V_0 R^2 \Delta \eta$. The angle $\delta_m$ is the opening angle under which the catalytic 
cap is seen from the center of the particle when the cap touches the interface 
(Fig. \ref{fig:stab-janus}(b)) and thus measures its size. (b) Approximate expression for 
$\Lambda$ as a function of $-\Pi > 0$ for $\delta_m = 0,\pi/4$, and $\pi/3$. Although 
the value $\delta_m = 0$ is unphysical, because in this case the catalytic cap reduces to 
a point, the corresponding curve represents the limiting case for which $\Lambda$ 
attains its maximal values.
} 
\label{fig:lambda}
\end{figure}

From Eqs. (\ref{def:bar-v}) and (\ref{def:distr-prob}), after disregarding corrections 
to $U_x$ of order $\epsilon = \frac{\Delta\eta}{\eta_0}$ (Eq.(\ref{Ux_expl})), one 
obtains
\begin{equation}
\label{eq:Lamb_calc}
\Lambda := \dfrac{\lambda_i}{\lambda_b} \simeq \dfrac{\tau_i}{\tau_b}
\dfrac{\bar{v}_i}{\bar{v}_b} \simeq 
\nu_0 \int \limits_{\delta_m}^{\pi/2} 
\dfrac{e^{-\beta\Phi(\delta)}}{Z} \cos(\delta)d\delta 
\end{equation}
By inserting $Z = \int \limits_{\delta_m}^{\pi/2} e^{-\beta\Phi(\delta)} d\delta$ and 
Eq. (\ref{def:pot-PHI}) into Eq. (\ref{eq:Lamb_calc}), the dependences of $\Lambda$ on 
$\delta_m$ and on $\Pi$ can be calculated. For the choice $\nu_0 = 3/2$ these are 
shown in Figs. \ref{fig:lambda}(a) and (b), respectively. (We recall that we have 
focused on the case of the catalytic cap exposed to the \textit{upper} phase.) 
As shown in Fig. \ref{fig:lambda}, for sufficiently negative values of $\Pi$
the persistence length of a Janus particle moving at a liquid-fluid interface 
may be \textit{larger} than the one in the corresponding bulk case, i.e., 
$\Lambda > 1$. According to the discussion in the previous paragraphs, the case of 
$\Pi < 0$ with the catalytic cap exposed to the upper phase corresponds to 
either $\Delta \eta > 0$ and repulsive interactions or $\Delta \eta < 0$ and attractive 
interactions, i.e., the cases for which sustained motility emerges (see 
Fig. \ref{fig:stability-condition}(b)). On the other hand, we have noted that for 
a given type of interactions (i.e., a given sign of $V_0$) and a given 
viscosity contrast $\Delta \eta$, the amplitude $\Pi$ of the potential $\Phi$
changes sign if the catalytic cap is exposed to the lower phase, i.e., 
for $\delta > \pi$, relative to the the case of the catalytic cap being 
exposed to the upper phase. Therefore, these corresponding dependences on 
$\delta_m$ and on $\Pi$ are given by the curves in Fig. \ref{fig:lambda}(a) 
and (b) but with the opposite sign of $\Pi$ and with $\delta_m \to -\delta_m$. 
Consequently, one infers that in this case one has $\Lambda > 1$ for sufficiently 
large positive values of $\Pi$. Thus also in the case that the cap is immersed in the 
lower phase the persistence length at the interface may be enhanced relative to 
the bulk one for those states in which sustained motility emerges. In summary, 
this implies that in all cases of sustained motility (i.e., the system 
corresponds to any of the cases shown in Fig. \ref{fig:stability-condition}(b)) 
the particle trapped at the interface exhibits an enhanced persistence length for 
sufficiently large values of $|\Pi|$.

Figure \ref{fig:lambda}(b) shows the dependence of $\Lambda$ on $\Pi$ for the 
case in which the catalytic cap is exposed to the upper phase. Interestingly, 
$\Lambda$ saturates at negative values of $\Pi$ with large $|\Pi|$. The saturation occurs 
at larger values of $|\Pi|$ upon increasing $\delta_m$. Concerning the magnitude of 
$\Pi$ at which $\Lambda$ starts to saturate, we recall that 
$|\Pi|=\frac{1}{2 \sqrt{3 \pi}}\frac{\Delta \eta}{\eta_0} Pe_0$. Therefore, for 
$\frac{\Delta\eta}{\eta_0}\rightarrow 0$ the onset of saturation at $|\Pi|\approx 5$ 
requires, even for very small caps, i.e., $\delta_m \to 0$, P\'eclet numbers 
$Pe_0 \simeq 30 \times (\Delta \eta/\eta_0)^{-1}$ much larger than the typical values 
$Pe_0 \simeq 10$ for Janus particles in a homogeneous bulk fluid. If, however, the 
viscosity contrast is high, the required corresponding $Pe_0$ numbers are 
significantly smaller. For example, for the water-air interface the viscosity of the air is 
negligible so that $\Delta\eta = - 2 \eta_0$ which implies $|\Pi| 
\simeq \frac{1}{3} Pe_0$. In such a situation, as well as for other liquid-fluid interfaces 
characterized by high viscosity contrasts, large values of $|\Pi|$ are encountered 
already for typical values of $Pe_0$ and the persistence length at the interface may 
be enhanced relative to its bulk value, i.e., $\Lambda > 1$. As shown in 
Fig. \ref{fig:lambda}, this effect is particularly pronounced for small catalytic caps 
(i.e., $\delta_m$ small).

While the persistence length $\lambda$ characterizes the active motion of a particle 
at time scales shorter than the characteristic rotational diffusion time $\tau$, at time 
scales much larger than $\tau$ the motion of the particle crosses over to diffusion with 
an effective diffusion constant \cite{GolestanianPRL2007}:
\begin{equation}
 D_{eff} = D_{tr} + \dfrac{\lambda^2}{\tau} : = D_{tr} + \delta D\,,
\label{eq:D_def}
\end{equation}
where $D_{tr}$ is the translational diffusion constant of the particle in the absence 
of activity and $\delta D = \lambda^2/\tau$ is the activity-induced enhancement of the 
diffusion constant. Equation (\ref{eq:D_def}) allows us to compare the enhancement 
$\delta D^{(i)}$ for the Janus particle trapped at, and moving along, the \textit{interface} 
to the one, $\delta D^{(b)}$, which holds for the same active Janus particle moving 
in a \textit{bulk} fluid:
\begin{equation}
 \dfrac{\delta D^{(i)}}{\delta D^{(b)}} = \left(\dfrac{\lambda_i}{\lambda_b}\right)^2 
\dfrac{\tau_b}{\tau_i} \simeq \dfrac{1}{\nu_0} \Lambda^2 \,.
\label{eq:deltaD}
\end{equation}
Therefore, according to the values of $\Lambda$ shown in Fig. \ref{fig:lambda}, for 
$\nu_0 = 3/2$ the enhancement of the diffusion constant due to the activity of a Janus 
particle trapped at a liquid-fluid interface can be up to $1.5$ times larger than the 
enhancement observed in a homogeneous bulk fluid. Finally, we note that 
for a particle trapped at the interface the activity induced contribution $\delta 
D^{(i)}$ can become much larger than the passive translational diffusion constant $D_P$ in 
bulk fluid if $({\bar v}_i^2 \tau_i^2)/ (\tau_i D_P) \gg 1$ (see the definition of 
$\delta D^{(i)}$ above and Eq. (\ref{def:lambda})). By using 
${\bar v}_i = \Lambda {\bar v}_b/\nu_0$ (Eq. (\ref{eq:Lamb_calc})), 
$\bar{v}_b=V_0/\sqrt{3\pi}$, $Pe_0 = |V_0| R/D_P$, and $\tau_i/\tau_b = \nu_0$, and by 
taking $\tau_b = 1/D_P^{(rot)}$, where $D_P^{(rot)} = 4 D_P/(3 R^2)$ is the rotational 
diffusion constant of the particle in a homogeneous bulk fluid \cite{GolestanianPRL2007}, 
for $\nu_0 = 3/2$ and by using Eqs.~(\ref{eq:Lamb_calc})-(\ref{eq:deltaD}) the 
condition $\delta D^{(i)}/D_P \gg 1$ translates into the condition $Pe_0 \gg 4.3/\Lambda$ 
for the P{\'eclet} number of the particle.

\section{Summary and conclusions}

We have studied the behavior of a chemically active Janus particle trapped at a liquid-
fluid interface, under the assumptions that the activity of the particle does not affect 
the surface tension of the interface and that the interface can be assumed to be flat. If 
particles are moving in such a set-up (Fig.~\ref{fig:stab-janus}), a coupling between 
rotation and translation arises due to the viscosity contrast $\Delta \eta$ between the 
two adjacent fluids. Assuming that the particles are axisymmetric, and that both fluid 
phases are homogeneous and isotropic, the motile state of the particles is characterized 
by their linear velocity $U_x$ in the plane of the interface and their angular velocity 
$\Omega_z$ about an axis perpendicular to the plane of motion spanned by the 
interface normal and the velocity. 

In Sec.~\ref{sec:phoretic_velocities} we have determined the linear and 
angular velocity $U_x$ and $\Omega_z$, respectively, by using the Lorentz 
reciprocal theorem~\cite{Lorentz_original}. Therein the stress-free interface is accounted 
for by imposing corresponding boundary conditions on the fluid flow in both phases and the 
fluids are taken to be quiescent far away from the particle. The result 
in Eq.~(\ref{rhs_rec_theo}) is valid for an arbitrary viscosity contrast 
$\Delta\eta$, including the limit of vanishing values of $\Delta\eta$ 
as well as the case that one of the two phases has a vanishing viscosity.

Determining $U_x$ and $\Omega_z$ via the reciprocal theorem requires to solve two 
independent auxiliary problems involving translation and rotation of a particle trapped at 
a liquid-fluid interface. In order to be able to obtain analytical solutions, we 
have considered neutrally buoyant particles exhibiting a contact angle of 
$\pi/2$ with the planar interface. Under these assumptions it is possible to exploit the 
available analytical solution for the stress exerted on the fluid by a particle which is 
translating without rotation~\cite{Pozrikidis2007}. The case of a particle rotating at the 
interface is more challenging because it requires to determine the fluid flow close to the 
three-phase contact line formed as the intersection of the interface and the particle 
surface. In order to circumvent the issue of the motion of the three-phase contact line 
and in order to gain analytical insight into the problem, we have assumed that the fluid 
slips along the particle only in a small region close to the three-phase contact line. 
Accordingly, we can consider the fluid flow on each part of the surface of the  particle 
to be \textit{de facto} equal to the one which a particle experiences in a corresponding 
homogeneous bulk fluid under a no-slip condition on its surface 
(Eq. (\ref{notrasl-rot-noslip})). The general expressions for $U_x$ and $\Omega_z$ 
(Eqs. (\ref{Ux}) and (\ref{Omz})) show that, for $\Delta\eta\neq 0$, $\Omega_z$ is 
nonzero. Therefore the motility of the particle along the interface is 
strongly affected  by the change in the orientation of the axis of the particle relative 
to the interface normal. Accordingly, the velocity of the particle along the 
interface can be either enhanced or reduced. 

In Sec.~\ref{sec:results} A we have established a diagram 
(Fig.~\ref{fig:stability-condition}) describing the situations for which $\Omega_z$ 
promotes orientations of the Janus particle axis to be parallel to the interface, 
hence enforcing the  motile state of the particle. In particular, for 
repulsive interactions between the particle and the self-generated solute (e.g., for 
catalytic platinum caps on polystyrene particles suspended in water-peroxide 
solutions\footnote{In this case the repulsive character of interaction is inferred 
from the experimentally observed motion away from the platinum cap and under the 
assumption that the mechanism of motion is self-diffusiophoresis and that only the 
oxygen production and the corresponding surface gradients of oxygen are relevant.}) 
we have found that the motile state is fostered if the catalytic cap is immersed into 
the more viscous phase, while the opposite conclusion holds for an attractive 
interaction. Therefore, by tuning the viscosity contrast $\Delta\eta$, one can control the 
motility of Janus particles trapped at liquid-fluid interfaces. 

In Sec.~\ref{sec:results} B these general consideration have been extended further by 
specifying model particles which allow one to analyze the density profiles 
of the reaction product. The expansion of these profiles in terms of spherical harmonics 
shows that only the amplitude $A_{1,0}$ of the largest wavelength mode affects $U_x$ and 
$\Omega_z$ (see Appendix~\ref{deriv-C1-C2}). Accordingly, for the model considered here, 
different systems characterized by diverse physical properties (such as the viscosity 
contrast $\Delta\eta$, the catalytic reaction, or the interaction between the reaction 
product and the particle) but exhibiting the same value of $A_{1,0}$ lead to the same 
values of $U_x$ and $\Omega_z$ (see Eqs.~(\ref{Ux_expl}) and (\ref{Omz_expl})). In 
particular we have found that both $U_x$ and $\Omega_z$ are proportional to the 
velocity scale $V_0$ (Eq.~(\ref{eq:def_V0})) which depends linearly on the prefactor $\mathcal{L}$ (see Eq.\ref{eq:phor-mob}), the reaction rate $Q$ per area, the inverse mean 
viscosity $\eta_0$ and the inverse diffusivity $D_0$ of the reaction product. The 
angular velocity experienced by the particle of radius $R$ is proportional to $V_0/R$ and 
to the viscosity contrast $\Delta\eta$. 

If the angular velocity promotes the alignment of the axis of the particle with the 
interface, the persistence length of the particle increases. In order to quantify this 
effect, in Sec.~\ref{sec:persitence length} we have proposed a factorization of the 
probability distribution (Eq.~(\ref{def:distr-prob})) for the orientation of 
the axis of the particle into an equilibrium and into a non-equilibrium distribution 
induced by the angular velocity and we have constructed an effective potential $\Phi$ 
(Eq.~(\ref{def:pot-PHI})) describing the latter. The strength $|\Pi|$ 
(Eq.~(\ref{def:pot-PHI})) of this potential is proportional to the bulk 
P\'eclet number of the particle, which is of the order of $10$, and therefore may lead to 
an increase of the persistence length of a trapped active particle relative to its value 
in the bulk fluid. Figure \ref{fig:lambda}(b) shows that this enhancement increases with 
$|\Pi|$ as well as upon decreasing the size of the catalytic cap (which allows for smaller 
values of $\delta_m$ (see Figs. \ref{fig:stab-janus}(b) and \ref{fig:lambda}(a)). At
long timescales the motion of active Janus particles is characterized by an 
effective diffusion coefficient (see Eq. (\ref{eq:D_def})). Concerning this regime 
our results predict that $\Delta\eta$ as well as $\Pi$ control the enhancement of the 
effective diffusion coefficient. In particular, by using Eq. (\ref{eq:deltaD}) and the 
data in Fig.~\ref{fig:lambda}, we have found that the presence of the interface can 
almost double the activity induced enhancement of the diffusion coefficient compared with 
the one in a homogeneous bulk fluid.

In sum we have obtained the following main results:
\begin{itemize}
 \item{Within a minimalistic model of active Janus particles trapped at a 
liquid-fluid interface, we have characterized their dynamics and have shown that 
their motility is strongly affected by the angular velocity induced on the particle due to 
the viscosity contrast $\Delta \eta$ between the adjacent fluids.}
 \item{We have shown that the rotation-translation coupling induced by 
 $\Delta\eta$ can affect experimentally observable quantities such as the persistence 
length and the effective diffusion coefficient of active Janus particles  trapped at  
liquid-fluid interfaces. In particular, the behavior described by our model is in 
agreement with recently reported, corresponding experimental observations of 
increased persistence lengths for chemically active Janus particles at water-air 
interfaces \cite{Stocco}, and it sheds light on the proposition of an alternative 
explanation for the observed phenomenon.}
 \item{Since the viscosity contrast $\Delta \eta$ can control the performance of active 
particles moving at liquid-fluid interfaces, we suggest that it can be relevant also for 
the onset of instabilities of thin films covered by active particles~\cite{Stark2014}.}
\end{itemize}

Finally, we mention a few interesting extensions of the present study. Relaxing some of 
the simplifying assumptions employed here might shed light on alternative means to 
control active particles motility at liquid-fluid interfaces. In this respect we 
recall that we have assumed that the contact angle of the particle with the interface is 
$\pi/2$, and that pinning of the three-phase contact line is absent. Concerning the 
contact angle, we expect particles with a contact angle unequal $\pi/2$ to experience 
extra torques due to the offset of their center of mass from the plane of the interface. 
A similar scenario has been reported for particles which are pulled, without rotating, 
under the action of suitably distributed external forces and torques 
\cite{Pozrikidis2007}. On the other hand, pinning of the three-phase contact line 
might affect the effective rotational diffusion and, possibly, suppress it, as shown 
recently for an equilibrium system~\cite{Stocco2}. Therefore we expect that for 
active Janus particles trapped at liquid-fluid interfaces the pinning of the three-phase 
contact line can enhance the persistence length, and therefore the effective 
diffusion, as argued in Ref.~\cite{Stocco}, too.

\section*{Acknowledgments}
P.M. thanks Dr. Antonio Stocco for useful discussions and for providing the 
preprint version of Ref.~\cite{Stocco}.

\appendix

\section{Forces and torques}

Here we present the steps of the derivations leading to Eqs. (\ref{eq:recipr-prob-1}) 
and (\ref{eq:recipr-prob-2}). In order to simplify the calculations, here it is convenient 
to translate the origin $O$ of the unprimed coordinate system (fixed in space, see 
Fig. \ref{fig:stab-janus}(b)) to the center $C$ of the moving particle and to use 
spherical coordinates $(r,\theta,\phi)$, which are defined as usual: $x = r \sin(\theta) 
\cos(\phi)$,  $y = r \sin(\theta) \sin(\phi)$, and $z = r \cos(\theta)$. (Note that these 
are defined in the unprimed coordinate system which, although exhibiting here the 
same origin as the primed (co-moving) one, has different orientations of the 
axes as compared with the primed one. The unprimed coordinate system offers a less 
cumbersome parametrization of the location of the interface as compared with the primed 
coordinate system.) We start with deriving Eq. (\ref{eq:recipr-prob-1}).

By using that for a translation (only) with velocity 
$\hat{\mathbf{U}}=\hat{U}_x\mathbf{e}_x$ one has \cite{Pozrikidis2007}
\begin{equation}
\left.\mathbf{n} \cdot \hat{\boldsymbol{\sigma}} \right|_{\Sigma_p}= -\frac{3}{2R} 
\eta(\mathbf{{r}}_p) \hat{\mathbf{U}}\,
\end{equation}
on the surface $\Sigma_p$ of the particle. By noting that
\begin{equation}
\hat{F}_x=\int_{\Sigma_p}(\mathbf{n} \cdot \hat{\boldsymbol{\sigma}})_x \, dS
\end{equation}
is the $x$ component of the integral over the surface of the normal pressure tensor 
as given by the above expression, using Eq. (\ref{eq:def-visc-1}), and assuming a sharp 
interface ($\xi \to 0$ so that $\eta(\mathbf{r}_p) = \eta(\phi)$ 
with $\eta(\phi) = \eta_1$ for $0 < \phi < \pi$, i.e., $y > 0$, while $\eta(\phi) = \eta_2$ 
for $\pi < \phi < 2\pi$, i.e., $y < 0$) one obtains
\begin{equation}
F_{x} =-\dfrac{3}{2R} \hat{U}_{x} R^{2} \int\limits_{0}^{\pi} d \theta \, 
\sin\left(\theta \right) 
\int\limits_{0}^{2\pi} d \phi \, \eta\left(\phi\right) = 
-6 \pi \eta_{0} R \hat{U}_{x}\,,
\end{equation}
which agrees with Eq. (\ref{F1x}). The torque is defined as
\begin{equation}
\mathbf{L} = \int_{\Sigma_p}(\mathbf{r}_p-\mathbf{r}_C) 
\times \hat{\boldsymbol{\sigma}} \cdot \mathbf{n}\,dS = R \int\limits_{\Sigma_p}^{} 
\mathbf{n} \times \hat{\boldsymbol{\sigma}} \cdot \mathbf{n} \, dS 
= -\dfrac{3}{2} \int\limits_{\Sigma_p}^{} \eta(\mathbf{{r}}) \, \mathbf{n} \times 
\hat{\mathbf{U}}\, dS\,,
\end{equation}
where we have used $\mathbf{r}_p-\mathbf{r}_C = R\,\mathbf{n}$ 
(see Fig. \ref{fig:stab-janus}). With the above 
choice of the coordinate system the torque is along the $z$-direction. Since 
$\left( \mathbf{n} \times \hat{\mathbf{U}} \right)_{z} = n_{x} \hat{U}_{y} - 
n_{y}\hat{U}_{x}$ and $\hat{U}_{y} = 0$ (due to the choice of the dual problem 
``1''), with $n_y = \sin(\theta) \sin(\phi)$ one obtains
\begin{equation}
L_{z}= \dfrac{3}{2} R^{2}\hat{U}_{x} 
\int\limits_{0}^{\pi} d\theta \,\sin^{2} \left(\theta \right)
\int\limits_{0}^{2\pi} d \phi \sin\left(\phi \right) \eta\left(\phi \right) = 
\dfrac{3}{2} \pi R^{2} \Delta\eta \hat{U}_{x}\,,
\end{equation}
which agrees with Eq. (\ref{L1z}).

Concerning the derivation of Eq. (\ref{eq:recipr-prob-2}) we start from the relation
(compare Eq. (\ref{notrasl-rot-noslip}) with $\hat{\mathbf{\Omega}} = \Omega_z 
\mathbf{e}_z$)
\begin{equation}
\left.\mathbf{n} \cdot \hat{\boldsymbol{\sigma}} \right|_{\Sigma_p}= 
- 3 \eta(\mathbf{r}_p) \,\hat{\mathbf{\Omega}} \times \mathbf{n}\,;
\end{equation}
we note that $\left( \hat{\mathbf{\Omega}} \times \mathbf{n} \right)_{x} = 
\hat{\Omega}_{y} n_{z}-\hat{\Omega}_{z} n_{y}$ and recall that by definition of
the dual problem ``2'' only the component $\hat{\Omega}_{z}$ of the angular velocity 
is non-zero. Thus we obtain 
\begin{equation}
F_{x} = 3 \hat{\Omega}_{z} R^{2} \int\limits_{0}^{\pi} d\theta 
\sin^{2}\left(\theta \right) 
\int\limits_{0}^{2\pi} d\phi \, \sin\left(\phi \right) \eta\left(\phi \right) d\phi
= 3 \pi R^{2} \Delta\eta \hat{\Omega}_{z}\,,
\end{equation}
which agrees with Eq. (\ref{F2x}). For the torque one has, with 
$\mathbf{r}_p-\mathbf{r}_C = R \mathbf{n}$,
\begin{eqnarray}
\label{eq:Lz_expr}
L_{z} = R \int\limits_{\Sigma_p}^{} \left[\mathbf{n} \times \hat{\boldsymbol{\sigma}} 
\cdot \mathbf{n} \right]_{z} \, dS &=& - 3 R \int\limits_{\Sigma_p}^{} 
\eta(\textbf{{r}}) \left[\mathbf{n} \times \left(\hat{\mathbf{\Omega}} \times 
\mathbf{n} \right)\right]_{z}\, dS = 
-3 R \int\limits_{\Sigma_p}^{} \eta(\textbf{{r}}) \hat{\Omega}_{z} \sin^2(\theta) 
\, dS \nonumber\\
& = & -3 R^3 \hat{\Omega}_{z} \int\limits_{0}^{\pi} d \theta \sin^{3} 
\left(\theta\right) \int\limits_{0}^{2 \pi} d\phi \,\eta(\phi) = -8 \pi \eta_{0} R^{3} 
\hat{\Omega}_{z} \,, 
\end{eqnarray}
which agrees with Eq. (\ref{L2z}) and where the following relations have been used:
\begin{equation}
 \hat{\mathbf{\Omega}} \times \mathbf{n} = -\left(
\begin{array}{c}
n_{y}\hat{\Omega}_{z}-n_{z}\hat{\Omega}_{y}\\
n_{z}\hat{\Omega}_{x}-n_{x}\hat{\Omega}_{z}\\
n_{x}\hat{\Omega}_{y}-n_{y}\hat{\Omega}_{x}
\end{array}
\right)
\end{equation}
and
\begin{equation}
\label{A10}
\left[\mathbf{n} \times \left( \hat{\mathbf{\Omega}} \times \mathbf{n} \right) \right]_{z}
= n_{x} n_{y} \hat{\Omega}_{x} + n_{x}^{2} \hat{\Omega}_{z} +  n_{y}^{2} \hat{\Omega}_{z} 
+ n_{y} n_{z} \hat{\Omega}_{y}
= \hat{\Omega}_{z} \left( n_{x}^{2} + n_{y}^{2} \right) 
= \hat{\Omega}_{z} \sin^{2}\left(\theta\right).
\end{equation}
In Eq. (\ref{A10}) we have used, according to the definition of the dual problem 
``2'',  $\hat{\Omega}_{x} = \hat{\Omega}_{y} = 0$.

\section{Diffusiophoretic slip \label{deriv-C1-C2}}

We start the derivation of Eqs. (\ref{eq:C1B}) and (\ref{eq:C2B}) by employing spherical 
coordinates $(r',\theta',\phi')$:
\begin{equation}
x' = r' \sin\left(\theta'\right)\cos\left(\phi'\right)\,,~
y' = r' \sin\left(\theta'\right)\sin\left(\phi'\right)\,,~
z' = r' \cos\left(\theta'\right)\,
\label{sph-coord}
\end{equation}in the co-moving (primed) coordinate system (see Fig. 
\ref{fig:stab-janus}(b)) with $\rho(\mathbf{r'}_p) = 
\rho(R,\theta',\phi')$; in the following, for reasons of shorter notations we 
shall not indicate explicitly the dependence on $R$. The density can be expressed as a 
series expansion in terms of the spherical harmonics $Y_{\ell\,m} (\theta',\phi')$:
\begin{equation}
\rho(\theta',\phi') = \sum\limits_{\ell = 0}^\infty  
\sum\limits_{m = -\ell}^{m = \ell} A_{\ell,m}\, Y_{\ell\,m} 
(\theta',\phi') \,,
\label{exp:rho_sph_harm}
\end{equation}
where \cite{Jackson_book}
\begin{equation}
 \label{sph_harm}
Y_{\ell\,m} (\theta',\phi') =  \alpha_{\ell,m} P_{\ell}^m (\cos\theta') 
e^{i m \phi'}\,,~i = \sqrt{-1}\,,
\end{equation}
and
\begin{equation}
P_{\ell}^m (x) = \dfrac{(-1)^m}{2^\ell \ell!} (1-x^2)^{m/2} 
\dfrac{d^{\ell+m}}{dx^{\ell+m}}  (x^2-1)^\ell\,,~ \ell \geq 0\,,~|m| \leq \ell
\end{equation}
is the associated Legendre polynomial of degree $\ell$ and order $m$ with
\begin{equation}
 \alpha_{\ell,m} = \sqrt{\dfrac{2 \ell + 1}{4 \pi}\dfrac{(\ell-m)!}{(\ell+m)!}}
\label{eq:def-alph}
\end{equation}
as a normalization constant.
Before proceeding, we list a few relations (obtained straightforwardly from the 
corresponding definitions) satisfied by $Y_{\ell\,m}$, $P_{\ell}^m$, and 
$\alpha_{\ell,m}$, which will be needed below\footnote{For completeness, we note 
that (i) the density must take real values, i.e., $\rho^*(\theta,\phi) = 
\rho(\theta,\phi)$, and (ii) the system is symmetric with respect to the plane $y'z'$ 
(see Fig. \ref{fig:stab-janus}(b)). Thus the solute distribution must be invariant 
with respect to the transformation $\phi' = \pi/2 - \epsilon \to \phi' = \pi/2 + 
\epsilon$, i.e., $\rho(\theta,\pi/2 - \epsilon) = \rho(\theta,\pi/2 + \epsilon)$. 
These properties require that the coefficients $A_{\ell,m}$ obey the relations 
$A^{^*}_{\ell,-m} = (-1)^m A_{\ell,m}$ and $A_{\ell,-m} = A_{\ell,m}$. This implies that 
all even coefficients, and in particular $A_{1,0}$, are real numbers.} :
\begin{eqnarray}
&&Y_{0\,0} = \dfrac{1}{\sqrt{4 \pi}}\,, \label{Y_00} \\
&&P_{\ell}^{-m} = (-1)^m \,\dfrac{(\ell-m)!}{(\ell+m)!} \, P_{\ell}^m\,, \label{Pl_min}\\
&&Y_{\ell(-m)} = (-1)^m \, Y^*_{\ell\,m} \,, \label{Yl_min} \\
&&\alpha_{\ell,-m} = \dfrac{(\ell + m)!}{(\ell - m)!} \, \alpha_{\ell,m}\,, 
\label{alfl_min} \\
&&\int\limits_0^{2 \pi} d\phi' \int\limits_0^{\pi} d\theta'\, \sin\theta \,
{Y^*_{\ell'\,m'}}(\theta',\phi') \, Y_{\ell \,m}(\theta',\phi') = 
\delta_{\ell,\ell'}\delta_{m,m'}\,, \label{normal} \\
&& \dfrac{\partial Y_{\ell\,m}}{\partial \phi} = i \, m \,Y_{\ell\,m}\,, 
\label{phi_der}
\end{eqnarray}
and
\begin{eqnarray}
&& \sin\theta' \dfrac{\partial Y_{\ell\,m}}{\partial\theta'} = \ell\, 
\sqrt{\dfrac{(\ell+1)^2 - m^2}{(2 \ell+1) (2 \ell+3)}} \, Y_{(\ell+1) \,m} - 
(\ell+1)\, 
\sqrt{\dfrac{\ell^2 - m^2}{(2 \ell-1) (2 \ell+1)}} \, Y_{(\ell-1) \,m} \nonumber\\
&&\hspace*{.6in}:= a_{\ell,m} Y_{(\ell+1) \,m} + b_{\ell,m} Y_{(\ell-1)\,m}\,, 
\label{theta_der}
\end{eqnarray}
where $^*$ indicates the complex conjugate quantity.

From the definition of the phoretic slip $\mathbf{v}(\mathbf{r}'_p)$ 
(Eq. (\ref{def:slip_vel})) one obtains 
\begin{equation}
\eta(\mathbf{r'}_p) v_{\theta'} = - \dfrac{\mathcal{L}}{\beta R} 
\sum\limits_{\ell = 0}^\infty  \sum\limits_{m = -\ell}^{m = \ell} A_{\ell,m}\, 
\dfrac{\partial Y_{\ell\,m}(\theta',\phi')} {\partial {\theta'}} \label{eq:B_vel_theta}
\end{equation}
and
\begin{equation}
\eta(\mathbf{r}'_p) v_{\phi'} = - \dfrac{\mathcal{L}}{\beta R} 
\dfrac{1}{\sin\theta'} \sum\limits_{\ell = 0}^\infty  
\sum\limits_{m = -\ell}^{m = \ell} A_{\ell,m}\, 
\dfrac{\partial Y_{\ell\,m}(\theta',\phi')} {\partial {\phi'}} \,.\label{eq:B_vel_phi}
\end{equation}
Noting that the unit vectors $\mathbf{e}_{r'}$, $\mathbf{e}_{\theta'}$, and 
$\mathbf{e}_{\phi'}$ are given by 
\begin{eqnarray}
\label{unit_vec}
\mathbf{e}_{r'} &=& \sin\theta' \cos\phi' \mathbf{e}_{x'} + 
\sin\theta' \sin\phi' \mathbf{e}_{y'} + \cos\theta' \mathbf{e}_{z'} \,,\nonumber\\
\mathbf{e}_{\theta'} &=& \cos\theta' \cos\phi' \mathbf{e}_{x'} + 
\cos\theta' \sin\phi' \mathbf{e}_{y'} - \sin\theta' \mathbf{e}_{z'}\,,\\
\mathbf{e}_{\phi'} &=& -\sin\phi' \mathbf{e}_{x'} + \cos\phi' \mathbf{e}_{y'} \nonumber
\end{eqnarray}
and using geometry (see Fig. \ref{fig:stab-janus}(b)) one obtains
\begin{equation}
\label{slip_x}
v_x = v_{z'} \cos \delta = (\mathbf{v} \cdot \mathbf{e}_{z'}) \cos\delta = 
(v_{\theta'} \mathbf{e}_{\theta'} \cdot \mathbf{e}_{z'} + 
v_{\phi'}  \mathbf{e}_{\phi'} \cdot \mathbf{e}_{z'}) \cos\delta 
\stackrel{(\ref{unit_vec})}{=} - v_{\theta'} \sin\theta' \cos\delta\,;
\end{equation}
according to Eq. (\ref{def:slip_vel}) $v_{r'} = 0$. Therefore
\begin{eqnarray}
C_{1} &:=& \dfrac{3}{2 R} \int\limits_{\Sigma_p}^{} dS\, \eta(\mathbf{r}'_p) \, v_x 
\nonumber\\
&\underset{(\ref{eq:B_vel_theta})}{\overset{(\ref{slip_x})}{=}}&
\dfrac{3}{2} \dfrac{\mathcal{L} \cos \delta}{\beta} \int\limits_0^{2 \pi} d \phi' 
\int\limits_0^{\pi} d\theta'\, \sin\theta' 
\sum\limits_{\ell = 0}^\infty  \sum\limits_{m = -\ell}^{m = \ell} A_{\ell,m}\, 
\sin\theta' \, 
\dfrac{\partial Y_{\ell\,m}(\theta',\phi')} {\partial {\theta'}} \nonumber\\
&\underset{(\ref{Y_00})}{\overset{(\ref{theta_der})}{=}}&
\dfrac{\mathcal{L} \cos \delta}{\beta} 
\sum\limits_{\ell = 0}^\infty  \sum\limits_{m = -\ell}^{m = \ell} A_{\ell,m}\, 
\left[a_{\ell,m} 
\int\limits_0^{2 \pi} d \phi' \int\limits_0^{\pi} d\theta'\, \sin\theta' 
Y^*_{0\,0}(\theta',\phi') Y_{(\ell+1)\,m}(\theta',\phi') \right. \nonumber\\
&&\hspace*{1.5in} \left. + ~ b_{\ell,m} 
\int\limits_0^{2 \pi} d \phi' \int\limits_0^{\pi} d\theta'\, \sin\theta' 
Y^*_{0\,0}(\theta',\phi') Y_{(\ell-1)\,m}(\theta',\phi')
\right]\nonumber\\
&\stackrel{(\ref{normal})}{=}& 
3 \sqrt{\pi} \dfrac{\mathcal{L} \cos \delta}{\beta} 
\sum\limits_{\ell = 0}^\infty  \sum\limits_{m = -\ell}^{m = \ell} A_{\ell,m}\, 
\left(a_{\ell,m} \delta_{0,\ell+1}\delta_{0,m} + b_{\ell,m} \delta_{0,\ell-1} 
\delta_{0,m}\right) = 3 \sqrt{\pi} \dfrac{\mathcal{L} \cos \delta}{\beta} 
A_{1,0}\, b_{1,0}\nonumber\\
&\stackrel{(\ref{theta_der})}{=}& -2 \sqrt{3 \pi} \,\dfrac{\mathcal{L} \cos \delta}{\beta} 
A_{1,0}\,,
\end{eqnarray}
which agrees with Eq. (\ref{eq:C1B}).

We now proceed with the calculation of $C_2$. First, we note that 
$\mathbf{e}_z$ = $\mathbf{e}_{x'}$ (see Fig. \ref{fig:stab-janus}(b)), and therefore 
$(\mathbf{n} \times \mathbf{v})_{z} := (\mathbf{n} \times \mathbf{v}) \cdot \mathbf{e}_z 
= (\mathbf{n} \times \mathbf{v})_{x'}$, where $\mathbf{v} = 
\mathbf{v}(\mathbf{r}'_p)$. 
The latter but one expression is calculated as follows (note that in the primed 
coordinate system $\mathbf{n} = \mathbf{e}_{r'}$):
\begin{eqnarray}
\label{cross_prod}
(\mathbf{n} \times \mathbf{v})_{x'} 
&=& 
n_{y'} v_{z'} - n_{z'} v_{y'} 
= (\mathbf{e}_{r'} \cdot \mathbf{e}_{y'}) v_{z'} - (\mathbf{e}_{r'} \cdot \mathbf{e}_{z'}) 
\left[v_{\theta'} (\mathbf{e}_{\theta'} \cdot \mathbf{e}_{y'}) + 
v_{\phi'} (\mathbf{e}_{\phi'} \cdot \mathbf{e}_{y'})\right]\nonumber\\
&\underset{(\ref{unit_vec})}{\overset{(\ref{slip_x})}{=}}&
v_{\theta'} - \cos\theta' (\cos\theta'\sin\phi' v_{\theta'} + \cos\phi' v_{\phi'}) 
\nonumber\\
&=& - (\sin\phi' v_{\theta'} + \cos\theta' \cos\phi' v_{\phi'})\,.
\end{eqnarray}
Therefore $C_2$ takes the form (Eq. (\ref{eq:rec-ther-2}))
\begin{eqnarray}
 \label{def_I1_I2}
C_2 &:=& \dfrac{3}{R} \int\limits_{\Sigma_p}^{} d S' \,\eta(\mathbf{r}'_p) 
(\mathbf{n} \times \mathbf{v})_{z} \nonumber\\
&=& - 3 R \left(
\int\limits_0^{2 \pi} d \phi' \sin\phi' \int\limits_0^{\pi} d\theta'\, \sin\theta' 
\eta(\mathbf{r'}_p) v_{\theta'} + 
\int\limits_0^{2 \pi} d \phi' \cos\phi' \int\limits_0^{\pi} 
d\theta'\, \sin\theta' \cos\theta' \eta(\mathbf{r'}_p) v_{\phi'}\right) \nonumber\\
&=:& 3 \dfrac{\mathcal{L}}{\beta} \left(J_1 + J_2\right)\,.
\end{eqnarray}
The integrals $J_1$ and  $J_2$ are evaluated as follows. By introducing the 
notations
\begin{equation}
\label{Z_def}
Z_\ell^m(\theta') = \dfrac{d P_\ell^m(\cos\theta')}{d \theta'}\,
\end{equation}
and 
\begin{eqnarray}
\label{z_p_def}
z_{\ell,m} &=& \int\limits_0^{\pi} d\theta' \, \sin\theta' \, 
Z_\ell^m(\theta')\,,\nonumber\\
p_{\ell,m} &=& \int\limits_0^{\pi} d\theta' \, \cos\theta' \, 
P_\ell^m(\cos\theta')\,,
\end{eqnarray}
after observing that
\begin{eqnarray}
\label{z_min}
z_{\ell,-m} \stackrel{(\ref{Pl_min})}{=} (-1)^m \,\dfrac{(\ell-m)!}{(\ell+m)!} 
z_{\ell,m} \,,\nonumber\\
p_{\ell,-m} \stackrel{(\ref{Pl_min})}{=} (-1)^m \,\dfrac{(\ell-m)!}{(\ell+m)!} 
p_{\ell,m} \,,
\end{eqnarray}
and  with
\begin{equation}
 \label{eq:zm_pm}
 z_{\ell,m} ~
 \underset{(\ref{z_p_def})}{\overset{(\ref{Z_def})}{=}} ~
 \left. \left(\sin\theta' \, P_\ell^m(\cos\theta')\right)\right|_0^\pi - 
\int\limits_0^{\pi} d\theta' \, 
[d(\sin\theta')/d\theta'] \, P_\ell^m(\cos\theta') = - p_{\ell,m}\,
\end{equation}
one arrives at
\begin{eqnarray}
 J_1 &:=& 
\int\limits_0^{2 \pi} d \phi' \sin\phi' \int\limits_0^{\pi} d\theta'\, \sin\theta' 
\left(- \dfrac{\beta R}{\mathcal{L}} \, \eta(\mathbf{r'}_p) v_{\theta'} \right) \nonumber\\
&\underset{(\ref{Z_def})}{\overset{(\ref{eq:B_vel_theta})}{=}}&
\sum\limits_{\ell = 0}^\infty  \sum\limits_{m = -\ell}^{m = \ell} 
A_{\ell,m}\, \alpha_{\ell,m} \int\limits_0^{\pi} d\theta'\, \sin\theta' Z_\ell^m(\theta') 
\int\limits_0^{2 \pi} d \phi' \sin\phi' e^{i m \phi'} \nonumber\\
&=& 
- i \pi
\sum\limits_{\ell = 0}^\infty  \sum\limits_{m = -\ell}^{m = \ell} 
A_{\ell,m}\, \alpha_{\ell,m} z_{\ell,m} \left(\delta_{m,-1}-\delta_{m,1} \right) 
\nonumber\\
&\underset{(\ref{alfl_min})}{\overset{(\ref{z_min})}{=}}&
i \pi \sum\limits_{\ell = 1}^\infty (A_{\ell,-1} + A_{\ell,1}) 
\alpha_{\ell,1} z_{\ell,1} \nonumber
\end{eqnarray}
(note that $z_{0,1} = 0$ due to $|m| \leq \ell$) and
\begin{eqnarray}
 J_2 &:=& \int\limits_0^{2 \pi} d \phi' \cos\phi' \int\limits_0^{\pi} 
d\theta'\, \sin\theta' 
\cos \theta' \left(- \dfrac{\beta R}{\mathcal{L}} \, 
\eta(\mathbf{r'}_p) v_{\phi'} \right) 
\nonumber\\
&\underset{(\ref{eq:B_vel_phi})}{\overset{(\ref{phi_der})}{=}}&
\sum\limits_{\ell = 0}^\infty  \sum\limits_{m = -\ell}^{m = \ell} 
i \,m \, A_{\ell,m}\, \alpha_{\ell,m} 
\int\limits_0^{\pi} d\theta'\, \cos\theta' P_\ell^m(\cos\theta') 
\int\limits_0^{2 \pi} d \phi' \cos\phi' e^{i m \phi'} \nonumber\\
&=& 
i \,\pi\, \sum\limits_{\ell = 0}^\infty  \sum\limits_{m = -\ell}^{m = \ell} 
m A_{\ell,m}\, \alpha_{\ell,m} p_{\ell,m} \left(\delta_{m,-1} + \delta_{m,1} \right)
\nonumber\\
&\underset{(\ref{alfl_min})}{\overset{(\ref{z_min})}{=}}&
i \pi \sum\limits_{\ell = 1}^\infty (A_{\ell,-1} + A_{\ell,1}) \alpha_{\ell,1} 
p_{\ell,1}\,.\nonumber\\
&\stackrel{(\ref{eq:zm_pm})}{=}& 
- i \pi \sum\limits_{\ell = 1}^\infty (A_{\ell,-1} + A_{\ell,1}) \alpha_{\ell,1} 
z_{\ell,1} = -J_1\,.\nonumber
\end{eqnarray}
Therefore $C_2 = 3 \dfrac{\mathcal{L}}{\beta} (J_1 + J_2) = 0$, which verifies 
Eq. (\ref{eq:C2B}).

\bibliography{swimmers}
 
\end{document}